\documentclass[epj]{svjour}
\usepackage{color}
\usepackage{amsmath}
\usepackage{graphicx}
\usepackage{dcolumn}
\usepackage{bm}
\usepackage{hyperref}
\usepackage{longtable}
\usepackage{supertabular}
\usepackage{latexsym}
\usepackage{graphics}
\usepackage{epstopdf}
\begin{document}
\title{\large Bohr Hamiltonian with Hulth\'{e}n plus ring-shaped potential for triaxial nuclei with deformation-dependent mass term}
\author{A. Adahchour\inst{1}, S. Ait El Korchi\inst{1}, A. El Batoul\inst{1}, A. Lahbas\inst{1,2} \and M. Oulne\inst{1}
\thanks{\emph{corresponding author:}  oulne@uca.ac.ma}%
}                     
%
%
\institute{High Energy Physics and Astrophysics Laboratory, Department of Physics,  Faculty of Sciences Semlalia, Cadi Ayyad University,
P.O.B. 2390, Marrakesh 40000, Morocco. \and ESMaR, Department of Physics, Faculty of Sciences, Mohammed V University in Rabat, Morocco.}
\date{Received: date / Revised version: date}
%
\abstract{
In this work, we present a new version of the Bohr collective Hamiltonian for triaxial nuclei within Deformation-Dependent Mass formalism $(DDM)$ using the Hulth\'{e}n potential. We shall call the  developed model Z(5)-HD. Analytical expressions for energy spectra are derived by means of the recent version of the Asymptotic Iteration Method. The calculated numerical results of energies and $B(E2)$ transition rates are compared with the experimental data, and several theoretical results from  Z(5) model, the model Z(5)-H using the Hulth\'{e}n potential without $DDM$ formalism as well as theoretical predictions of Z(5)-DD model with Davidson potential using $DDM$ formalism. The obtained results show an overall agreement with experimental data and an important improvement in respect to the other models.
} 

\maketitle
\section{Introduction}
Since its introduction for the first time in semi-conductor physics \cite{Von1983}, the position dependent mass formalism has been applied in several works in many different fields of physics \cite{Bas1,Ser1,Saa1,Pla1,Alh1,Alh2,Ch6,Bo8,BO4,Soa,Sob,Ch8,Ch9}. Such a formalism has been adopted in nuclear physics for concrete reasons, namely: from the comparison of theoretical calculations with the experimental data, it has been pointed out that the mass tensor should be taken as a function of the collective coordinates \cite{BO4,Jol10,Jol11,Jol12,Jol13,Jol14,Erm10,Erm11}. So, based on these considerations, the Bohr Hamiltonian with a mass depending on collective variable can be elaborated for studying collective excited states in nuclei. These latter play an important role in the so-called shape phase transitions in nuclei which are of quantum type. Thus, several models considered as critical points of symmetries, namely: E(5) \cite{E5}, X(5) \cite{X5}, X(3) \cite{X3}, Z(5) \cite{Bo2}, Z(4) \cite{Z4} have been introduced.
\par In the present work, we will focus on the critical point symmetry Z(5) which
represents the transition from prolate axially symmetric SU(3) nuclei to oblate shape. We will consider a Bohr Hamiltonian with  Hulth\'{e}n potential including a mass parameter depending on the collective coordinate $\beta$. We have chosen this potential for its flatness. Indeed, it has been proved that as the considered potential is flat when $\beta$ increases as the calculated transition probabilities are more precise \cite{Ch7}. 
\par We will study the same isotopes that have been already treated in  \cite{BO4,Ch2,BO6} using respectively Davidson potential within the  Deformation Dependent Mass formalism (here called Z(5)-DD model), the Hulth\'{e}n potential without DDM (here called Z(5)-H model) and  Davidson potential without DDM (called  Z(5)-D model). Hence,  our aim is to study: \\ \\
\hspace*{0.3cm} 1) The effect of DDM on the energy spectra and $B(E2)$ transition probabilities of the isotopes $^{126,128,130,132,134}$Xe and $^{192,194,196}$Pt  using the same potential \\ \\
\hspace*{0.3cm} 2) The effect of the potential on the energy spectra and $B(E2)$ transition probabilities of the same isotopes taking into account the Deformation Dependent Mass term.
\par The structure of the present work is as follows: In Sections 2, 3 and 4, the theoretical background of the elaborated model Z(5)-HD is presented, namely: the $\beta$ part and the $\gamma$ part of the spectrum, the obtained analytical expressions for the energy levels by means of Asymptotic Iteration Method and the the total wave function. Section 5 contains the $B(E2)$ transition probabilities, while numerical results for energy spectra and $B(E2)$ are presented, discussed and compared with other results in section 6. Finally, section 7 contains a conclusion.        

\section{The Z(5)-HD model}
In the framework of Z(5), the original Bohr Hamiltonian \cite{B1} is
\begin{equation}
\begin{split}
		H_{B}=-\dfrac{\hbar^{2}}{2B} \bigg[\frac{1}{\beta^{4}}\frac{\partial}{\partial\beta}\beta^{4}\frac{\partial}{\partial\beta}+\frac{1}{\beta^{2}\sin3\gamma}\frac{\partial}{\partial\gamma}\sin3\gamma\frac{\partial}{\partial\gamma}  \\ -\frac{1}{4\beta^{2}}\sum^{3}_{k=1}\frac{Q_{k}^{2}}{\sin^{2}(\gamma-\frac{2\pi}{3}k)} \bigg]+ V(\beta,\gamma),
\end{split}
\label{HB1}
\end{equation}
where $B$ is the mass parameter, which is usually considered constant, $\beta$ and $\gamma$ are the usual collective coordinates ($\beta$ being a deformation coordinate measuring departure from spherical shape, and $\gamma$ being an angle measuring departure from axial symmetry), while $Q_{k}$ (k = 1, 2, 3) are the components of angular momentum in the intrinsic frame. 
\newline
By using a mass depending on the radial deformation coordinate $\beta$, 
\begin{equation}
   B(\beta)=\frac{B_{0}}{(f(\beta))^2},
\label{M1}
\end{equation}
where $B_{0}$ is the constant mass and $f(\beta)$ the deformation function, the Schr\"{o}dinger equation corresponding to the Hamiltonian \eqref{HB1} is given by \cite{BO4}
\begin{align} 
H\Psi(\beta,\gamma,\theta_{i})= &  \bigg[-\dfrac{1}{2}\dfrac{\sqrt f}{\beta^4}\dfrac{\partial}{\partial\beta}\beta^4 f\dfrac{\partial}{\partial\beta}\sqrt f    -\dfrac{f^2}{2\beta^2 \sin3\gamma} \notag \\
&   \dfrac{\partial}{\partial\gamma}\sin3\gamma \frac{\partial}{\partial\gamma}  
   +\; \frac{f^2}{8\beta^2}  \sum_{k=1,2,3}\frac {Q^{2}_{k}}{\sin^2(\gamma-\frac{2}{3}\pi k)}  \notag \\
&  +V_{eff} \bigg]\Psi(\beta,\gamma,\theta_{i}) =\varepsilon \Psi(\beta,\gamma,\theta_{i}), 
\label{Ha1}                                                                                   
\end{align}
where $\theta_{i}$ are the Euler angles and the reduced energies $\varepsilon$, reduced potential $v(\beta,\gamma)$, effective potential $V_{eff}(\beta,\gamma)$ are respectively \\
\hspace*{2cm} $\varepsilon=\frac{B_{0}}{\hbar^2 }\;E$ , 
\hspace*{0.2cm} $v(\beta,\gamma)=\frac{B_{0}}{\hbar^2 }\;V(\beta,\gamma)$ \\
\hspace*{2cm} $V_{eff}(\beta,\gamma)=v(\beta,\gamma)+\frac{1}{4}(1-\delta-\lambda)f\nabla^2 f \\ \hspace*{4.1cm} + \frac{1}{2}(\frac{1}{2}-\delta)(\frac{1}{2}-\lambda)(\nabla f)^2 $, \\
where $\delta$ and $\lambda$ are free parameters, as it was proved in \cite{Von1983} that the most general form of such a Hermitian Hamiltonian contains two free parameters (denoted by $\delta$ and $\lambda$ in the present work). These parameters came from the construction procedure of the kinetic energy term. In Refs \cite{BO4,Bag2005}, it has been shown that these parameters had no effect on the obtained results. The predictions for  theoretical spectra turn out to be independent of the choice made for these two free parameters. Also, in the present work (section 6), it will be seen that these parameters play practically no role. \\ 
The function $f(\beta)$ depends only on the radial coordinate $\beta$. So, only the $\beta$ part of the above equation is affected.
\section{Separable form of the Hamiltonian}
In order to achieve a separation of variables, we assume that the reduced potential $v(\beta,\gamma)$ depends on the variables $\beta$ and $\gamma$ and has the form \cite{Fo2,Fo3,Fo4,Bo3,Radu} \\
\begin{equation}
\hspace*{2cm}  v(\beta,\gamma)=u(\beta)+\frac{f^2}{\beta^2} \; w(\gamma),
\label{Pot1}
\end{equation}
with $w(\gamma)$ having a deep minimum at $\gamma$=$\frac{\pi}{6}$ and the wave functions have the form 
\begin{equation}
\hspace*{2cm} \Psi(\beta,\gamma,\theta_{i})=\xi(\beta) \; \Phi(\gamma,\theta_{i}).
\label{Fon1}
\end{equation}
The separation of variables gives
\begin{equation}
\begin{split} 
\bigg[-\dfrac{1}{2}\dfrac{\sqrt f}{\beta^4}\dfrac{\partial}{\partial\beta}\beta^4 f\dfrac{\partial}{\partial\beta}\sqrt f + \frac{f^2}{2\beta^2}\Lambda+ \frac{1}{4}(1-\delta-\lambda)f\nabla^2 f \\   +\;\frac{1}{2}(\frac{1}{2}-\delta)(\frac{1}{2}-\lambda)(\nabla f)^2 +u(\beta) \bigg] \; \xi(\beta)=\varepsilon \;\xi(\beta),
\end{split} 
\label{Bb}                                                                                   
\end{equation}
and
\begin{equation}
\begin{split} 
\bigg[-\dfrac{1}{\sin3\gamma}\dfrac{\partial}{\partial\gamma}\sin3\gamma \frac{\partial}{\partial\gamma}+ \frac{1}{4}  \sum_{k=1,2,3}\frac {Q^{2}_{k}}{\sin^2(\gamma-\frac{2}{3}\pi k)} \\ + \; w(\gamma) \bigg]\Phi(\gamma,\theta_{i}) =\Lambda \; \Phi(\gamma,\theta_{i}),
\end{split} 
\label{Bg}                                                                  
\end{equation}
where $\Lambda$ is the separation constant and equation \eqref{Bb} can be simplified by performing the derivatives
\begin{equation}
\begin{split} 
\frac{1}{2}f^2\xi^{''}+\bigg(ff^{'}+\frac{2f^2}{\beta}\bigg)\xi^{'}+\bigg[\frac{(f^{'})^2}{8}+\frac{ff^{''}}{4}+ \\   \frac{ff^{'}}{\beta}- \frac{f^2}{2\beta^2}\Lambda+ \varepsilon-v_{eff}\bigg]\xi=0,
\end{split} 
\label{Eqr1}
\end{equation}
with
\begin{equation}
v_{eff}=u(\beta)+\frac{1}{4}(1-\delta-\lambda)f(\frac{4f^{'}}{\beta}+f^{''})+\frac{1}{2}(\frac{1}{2}-\delta)(\frac{1}{2}-\lambda)(f^{'})^2.
\end{equation}
In the present work, we use the Hulth\'{e}n potential \cite{Hu1,Hu2} with a unit depth as in \cite{La1,Ma1}
\begin{equation}
u(\beta)=-\frac{1}{e^{\tau \beta}-1},
\label{Hu}
\end{equation}
where $\tau=\frac{1}{b}$ is a screening parameter and $b$ is the range of the potential. This potential has some properties, namely: it behaves as a short-range potential for small values of $\beta$ and decreases exponentially for very large values of $\beta$. By inserting the function $F(\beta)={\beta^2}\;\xi(\beta)$ in the radial equation \eqref{Eqr1}, one obtains
\begin{equation}
\begin{split} 
f^2 F^{''}+ 2ff^{'}F^{'}+\bigg(2\varepsilon-2(v_{eff}+\frac{f^2+\beta ff^{'}}{\beta^2}+   \\  \frac{f^2\Lambda}{2\beta^2}-  \frac{(f^{'})^2}{8}-\frac{ff^{''}}{4})\bigg)F=0 .
\end{split} 
\label{Eqr2}
\end{equation}
In order to make connection between our results and those obtained in Ref.\cite{Ch2}, we have replaced $2\varepsilon$ by $\epsilon$  and divided $u(\beta)$ by 2 in the above equation. So, one obtains
\begin{equation}
f^2 F^{''}+ 2ff^{'}F^{'}+(\epsilon-2u_{eff})F=0,
\label{Eqr3}
\end{equation}
where 
\begin{equation}
u_{eff}=v_{eff}+\frac{f^2+\beta ff^{'}}{\beta^2}+\frac{f^2\Lambda}{2\beta^2}-\frac{(f^{'})^2}{8}-\frac{ff^{''}}{4} .
\label{Pef}
\end{equation}
The special form for the deformation function is
\begin{equation}
f(\beta)=1+a\beta^2   ,\;\;\;  a<<1.
\label{Fdm}
\end{equation}
By inserting these forms for the potential and the deformation function in Eq. \eqref{Pef}, one gets
\begin{equation}
2u_{eff}=k_{1}\beta^2+k_{0}+\frac{k_{-1}}{\beta^2}-\frac{1}{e^{\tau\beta}-1},
\end{equation}
where 
\begin{align}
k_{1}&=a^2 \bigg(5(1-\delta-\lambda)+(1-2\delta)(1-2\lambda)+4+\Lambda \bigg), \nonumber \\
k_{0}&=a \bigg(5(1-\delta-\lambda)+7+2\Lambda \bigg),  \\
k_{-1}&= 2+\Lambda. \nonumber
\label{kkk}
\end{align}
Equation \eqref{Eqr3} becomes
\begingroup\small
\begin{equation}
f^2 F^{''}(\beta)+ 2ff^{'}F^{'}(\beta)+\bigg(\epsilon-k_{1}\beta^2-k_{0}-\frac{k_{-1}}{\beta^2}+\frac{1}{e^{\tau\beta}-1} \bigg)F(\beta)=0.
\label{Eqr4}
\end{equation}
\endgroup
To simplify equation \eqref{Eqr4}, we will proceed to a change of the function R($\beta$) by
\begin{equation}
F(\beta)=\frac{R(\beta)}{1+a\beta^2}.
\end{equation}
Thus, equation \eqref{Eqr4} becomes 
\begingroup\small
\begin{equation}
\begin{split}
R^{''}(\beta)+\bigg(\frac{\epsilon-k_{1}\beta^2-k_{0}}{(1+a\beta^2)^2}-\frac{2a}{1+a\beta^2}\;\;\;+ \\  \frac{1}{(1+a\beta^2)^2(e^{\tau\beta}-1)} -\frac{k_{-1}}{(1+a\beta^2)^{2}\;\beta^2}\bigg)R(\beta) =0.
\end{split}
\label{Eqr5}
\end{equation}
\endgroup
From this equation, if we set the deformation parameter $a = 0$, we recover the equation (7) of Ref. \cite{Ch2}. Because of the centrifugal potential and the form of the Hulth\'{e}n one, the Schr\"{o}dinger equation \eqref{Eqr5} cannot be solved analytically. So we will proceed to a rigorous approximation that allows to tackle this problem. For a small $\beta$ deformation, the centrifugal potential could be approximated by the following expression as in Refs. \cite{Ji1,Do1,So1}
\begin{equation}
\frac{1}{\beta^2}  \approx  \tau^2\frac{e^{-\tau\beta}}{(e^{-\tau\beta}-1)^2}.
\end{equation}
This approximation is also valid for small values of the
screening parameter $\tau$. By using the new variable $y=e^{-\tau\beta}$, we obtain
\begin{equation}
\begin{split} 
\frac{1}{e^{\tau\beta}-1}=\frac{y}{1-y}\;\; , \;\;\;\; \beta=\frac{1-y}{\tau \;\sqrt y } \;\;,\;\;\;  \\  1+a\beta^2=\frac{a(1-y)^2+\tau^2 \; y}{\tau^2 \; y}.
\end{split} 
\end{equation}
Rewriting equation \eqref{Eqr5} by using the new variable $y$, we get
\begingroup\small
\begin{equation}
\begin{split}
R^{''}(y)+\frac{1}{y}R^{'}(y)+\bigg[\frac{(\epsilon-k_{0})\tau^2+(2-y)k_{1}}{(ay^2+(\tau^2-2a)y+a)^2}- \\ \frac{k_{1}+2a}{y(ay^2+(\tau^2-2a)y+a)^{2}} +\frac{\tau^2\;y}{(1-y)(ay^2+(\tau^2-2a)y+a)^2} \\-   \frac{\tau^4k_{-1}\;y}{(1-y)^2(ay^2+(\tau^2-2a)y+a)^2}\bigg]R(y)=0.
\end{split}
\label{Eqr6}
\end{equation}
\endgroup
If $a = 0$, the dependence of the mass on the deformation is canceled, then we easily check that we recover the equation (9) of Ref. \cite{Ch2}. \\
The Schr\"{o}dinger equation \eqref{Eqr6} cannot yet be solved analytically because of some terms. Hence, in the absence of a rigorous solution to this equation,
we can use a further approximation. For a small deformation parameter $a$ ($a<<1$), and in view that $k_{1}$ is proportional to $a$ parameter,  as a first approximation, equation \eqref{Eqr6} becomes

\begin{equation}
\begin{split}
R^{''}(y)+ \frac{1}{y}R^{'}(y)+  \bigg[\frac{(\epsilon-k_{0})\tau^2+2k_{1}}{\tau^4y^2}
+\frac{1}{\tau^2 y(1-y)} \\- \frac{k_{-1}}{ y(1-y)^2}  \bigg]R(y)=0.
\end{split}
\label{Eqr7}
\end{equation}
In order to transform the above differential equation to a more compact one, we use the following variables
\begin{equation}
\mu^2=-\frac{(\epsilon-k_{0})\tau^2+2k_{1}}{\tau^4} , \;\;\;\; \nu=\frac{1}{2}(1+\sqrt{1+4k_{-1}}).
\label{par1}
\end{equation}
So, the differential equation \eqref{Eqr7} becomes
\begin{equation}
\begin{split}
R^{''}(y)+\frac{1}{y}R^{'}(y)-\left[\frac{\mu^2}{y^2}-\frac{1}{\tau^2y(1-y)}   
+\frac{\nu^2-\nu}{ y(1-y)^2}     \right]R(y)=0.
\end{split}
\label{Eqr8}
\end{equation}
To apply the asymptotic iteration method of Refs. \cite{Ci1,Ci2},  the reasonable physical wave function that  we propose is as follows
\begin{equation}
R(y)=y^{\mu}(1-y)^{\nu}\chi(y).
\label{par2}
\end{equation}
For this form of the radial wave function, Eq. \eqref{Eqr8} reads
\begin{equation}
\chi^{''}(y)=-\frac{\omega(y)}{\sigma(y)}\;\;\chi^{'}(y)-\frac{\kappa_{n}}{\sigma(y)}\;\;\chi(y),
\label{Echi}
\end{equation}
with
\begin{subequations}
\begin{align}
\omega(y)&=(2\mu+1)-(2\mu+2\nu+1)y,  \\
\sigma(y)&=y(1-y), \\
\kappa_{n}&=\frac{1}{\tau^2}-\nu(2\mu+\nu).   
\end{align}
\end{subequations}
Equation \eqref{Echi} leads us directly to the energy eigenvalues  using the new generalized formula \cite{Boz1} which replaced the iterative calculations in the original AIM formulation \cite{Ci3}.
\begin{equation}
\kappa_{n}=-n\;\omega^{'}(y)-\frac{n(n-1)}{2}\;\;\sigma^{''}(y).
\label{Ekappa}
\end{equation}
The above formulation gives the energy spectrum of the $\beta$ equation
\begin{equation}
\begin{split}
\epsilon_{n}=-\left(\frac{\tau^2(n+\frac{1}{2}+\sqrt{\frac{1}{4}+k_{-1}})^2-1}{2\tau(n+\frac{1}{2}+\sqrt{\frac{1}{4}+k_{-1}})}     \right)^2 - 2\;\frac{k_{1}}{\tau^2}+k_{0},
\end{split}
\label{En}
\end{equation}
where $n$ is the principal quantum number and $k_{-1}$, $k_0$ and $k_1$ are defined previously as a function of $\Lambda$, which represents the eigenvalues of the $\gamma$-vibrational plus rotational part of the Hamiltonian for triaxial nuclei. If we set the deformation parameter $a=0$, our energy  formula Eq. \eqref{En}  matches with that obtained in previous works  \cite{Ch2,Ik1,Ba2,Ag1}.\\
\\
For Eq.\eqref{Bg}, which represents the $\gamma$ variable, we use a new generalized potential proposed in \cite{Ch3} which is inspired by a ring-shaped potential
\begin{equation}
w(\gamma)=\frac{c+s\; \cos^{2}(3\gamma)}{\sin^{2}(3\gamma)},
\label{Potg}
\end{equation}
where $c$ and $s$ are free parameters. Inserting this form of the potential in equation \eqref{Bg}, we get
\begin{equation}
\begin{split} 
\bigg[-\dfrac{1}{\sin3\gamma}\dfrac{\partial}{\partial\gamma}\sin3\gamma \frac{\partial}{\partial\gamma}+ \frac{1}{4}  \sum_{k=1,2,3}\frac {Q^{2}_{k}}{\sin^2(\gamma-\frac{2}{3}\pi k)} \\ +\; \frac{c+s\; \cos^{2}(3\gamma)}{\sin^{2}(3\gamma)}\bigg]\Phi(\gamma,\theta_{i}) =\Lambda \; \Phi(\gamma,\theta_{i}).
\end{split} 
\label{Bg1}                                                                  
\end{equation}
Since the potential is minimal at $\gamma=\frac{\pi}{6}$, then the angular momentum term can be written as 
 \cite{Fo1,Bo1}
\begin{equation}
\frac{1}{4}  \sum_{k=1,2,3}\frac {Q^{2}_{k}}{\sin^2(\gamma-\frac{2}{3}\pi k)}\approx {\bf Q}^{2} -\frac{3}{4} Q_{1}^{2},
\label{Mo}                                                                  
\end{equation}
with ${\bf Q}^{2} = Q_{1}^{2}+Q_{2}^{2}+Q_{3}^{2}$ and the wave functions are given in the form 
\begin{equation}
\Phi(\gamma,\theta_{i})=\Gamma(\gamma) \; D_{M,\alpha}^{L}(\theta_{i}).
\label{F1}                                                                  
\end{equation}
Thus, the separation of variables leads to the following set of differential equations
\begin{equation}
\bigg[-\dfrac{1}{\sin3\gamma}\dfrac{\partial}{\partial\gamma}\sin3\gamma \frac{\partial}{\partial\gamma}+ \frac{c+s\; \cos^{2}(3\gamma)}{\sin^{2}(3\gamma)}\bigg]\; \Gamma(\gamma)=\Lambda^{'} \;\Gamma(\gamma),
\label{D1}                                                                  
\end{equation}

\begin{equation}
[Q^{2}-\frac{3}{4} Q_{1}^{2}]\; D_{M,\alpha}^{L}(\theta_{i})=\bar{\Lambda}\; D_{M,\alpha}^{L}(\theta_{i}).
\label{D2}                                                                  
\end{equation}
The resolution of the above  equation is carried out by Meyer-ter-Vehn \cite{B2,Me} with the results:
\begin{equation}
\bar{\Lambda} = L(L+1)-\frac{3}{4}\alpha^{2},
\label{La1}                                                                  
\end{equation}
\begingroup\small
\begin{equation}
D_{M,\alpha}^{L}(\theta_{i}) = \sqrt{\frac{2L+1}{16\pi^2(1+\delta_{\alpha,0})}} \bigg[ D_{M,\alpha}^{(L)}(\theta_{i})+(-1)^{L} D_{M,-\alpha}^{(L)}(\theta_{i}) \bigg],
\label{ED}                                                                  
\end{equation}
\endgroup
where $D(\theta_{i})$ denotes Wigner functions of the Euler angles $\theta_{i}(i = 1, 2, 3)$, $L$ is the total angular momentum quantum number, while $M$ and $\alpha$ are the quantum numbers of the projections of angular momentum on the laboratory fixed $z$-axis and the body-fixed $x^{'}$-axis, respectively.
\par In the study of triaxial nuclei, instead of the projection $\alpha$ of the angular momentum on the $x^{'}$-axis, we use the wobbling quantum number $n_{w}=L-\alpha$ \cite{B2,Me}. By replacing $\alpha$ by $L-n_{w}$ in Eq. \eqref{La1}, one obtains
\begin{equation}
\bar{\Lambda} = \frac{L(L+4)+3n_{w}(2L-n_{w})}{4}.
\label{La2}                                                                  
\end{equation}
For the sake of solving equation \eqref{D1} through the AIM, we introduce a new variable $z =$ cos$(3\gamma)$ and we propose the following ansatz for the eigenvectors $\Gamma(\gamma)$ 
\begin{equation}
\Gamma(z)=(1-z^2)^{\frac{1}{6}\sqrt{c+s}} \; \eta(z),
\label{Ga11}                                                                  
\end{equation}
leading to and
\begingroup\small
\begin{equation}
\eta^{''}(z)= - \frac{2(1+\frac{1}{3}\sqrt{c+s})z}{z^2-1}  \; \eta^{'}(z) \; - \frac{(3\sqrt{c+s}+c-\Lambda^{'})}{9(z^2-1)}\eta(z).
\label{eta2}                                                                  
\end{equation}
\endgroup
By using the generalized formula of AIM given in Eq. \eqref{Ekappa}, we derive the eigenvalues:
\begin{equation}
\Lambda^{'} = 9n_{\gamma}(n_{\gamma}+1)+3\sqrt{c+s}(2n_{\gamma}+1)+c,
\label{}                                                                  
\end{equation}
where $n_{\gamma}$ is the quantum number related to $\gamma$-excitations.\\
Finally, the analytical expression of  $\Lambda$, which represents the eigenvalues of the $\gamma$-vibrational plus rotational part of the Hamiltonian for triaxial nuclei is
\begin{equation}
\begin{split}
\Lambda = 9n_{\gamma}(n_{\gamma}+1)+3\sqrt{c+s}(2n_{\gamma}+1)+ \;c \; +  \\  \frac{L(L+4)+3n_{w}(2L-n_{w})}{4}.
\end{split}
\label{}                                                                  
\end{equation}
The solution of equation \eqref{eta2} gives the eigenfunctions  which are obtained in terms of Legendre polynomials.
\begin{equation}
\begin{split}
\eta(z)=N_{n_{\gamma}} \; (1-z^2)^{-\frac{1}{6}\sqrt{c+s}} \; P^{\frac{1}{3}\sqrt{c+s}}_{n_\gamma +\frac{1}{3}\sqrt{c+s}} (z),
\end{split}
\label{}                                                                  
\end{equation}
where $N_{n_{\gamma}}$ is a normalisation constant.\\
From equation \eqref{Ga11}, the $\gamma$ angular wave function for triaxial nuclei is given by
\begin{equation}
\begin{split}
\Gamma(\gamma)=N_{n_{\gamma}}  \; P^{\frac{1}{3}\sqrt{c+s}}_{n_\gamma +\frac{1}{3}\sqrt{c+s}} (\cos(3\gamma)),
\end{split}
\label{}                                                                  
\end{equation}
and the normalisation constant is obtained by using the normalisation condition
\begin{equation}
\begin{split}
\int_0^\frac{\pi}{3} \Gamma^2(\gamma) \;  |\sin(3\gamma)| \; d\gamma =1.
\end{split}
\label{}                                                                  
\end{equation}
By using the orthogonality relation of Legendre polynomials (See Ref. \cite{Int1}, Eq.(7.112), page 769), we obtain
\begin{equation}
\begin{split}
N_{n_{\gamma}}=\bigg[{\; \frac{(3(n_\gamma+\sqrt{c+s}/3)+\frac{3}{2}) \; (n_{\gamma} !) }{(2\sqrt{c+s}/3+n_\gamma)!}} \bigg]^\frac{1}{2}.   
\end{split}
\label{}                                                                  
\end{equation}
\section{Wave functions}
The total wave function has the form
\begin{align}
\Psi(\beta,\gamma,\theta_{i}) & =  \xi(\beta) \; \Phi(\gamma,\theta_{i})  \notag \\ 
&= \beta^{-2} R(\beta) \; \Gamma(\gamma) \; D^{L}_{M,\alpha_i}(\theta_{i}),
\label{AAA}
\end{align}
where $R(\beta)$ is the radial function corresponding to the eigenvectors of Eq. \eqref{Eqr7}, $\Gamma(\gamma)$ is the angular wave function of the $\gamma$-part given by Eq.\eqref{D1} and $D^{L}_{M,\alpha_i}(\theta_{i})$ are the eigenfunctions of the angular momentum given by Eq. \eqref{ED}. By using the parametrization given in Eqs. \eqref{par1}-\eqref{par2} and the general solution of Asymptotic Iteration Method (AIM), the solution of equation \eqref{Echi} is obtained as
\begin{equation}
\begin{split}
\chi(y)=N_n \;\; _{2}F_1( [-n,2\mu+2\nu+n], [2\mu+1],y ),
\end{split}
\label{}
\end{equation}
where $_{2}F_1$ are hyper-geometrical functions and $N_n$ is a normalization constant. So, 
by using the connection between the hyper-geometrical functions and Jacobi polynomials, we derive 
\begin{equation}
\begin{split}
\chi(y)=N_n \;\; \frac{\Gamma(2\mu+1)\Gamma(n+1)}{\Gamma(2\mu+n+1)}\;\; P_n(2\mu,2\nu-1)(y).
\end{split}
\label{}
\end{equation}
By proceeding to a change of variable $y$ in the function $R(y)$(Eq. \eqref{par2}) by $t=1-2y$, we obtain finally the following wave function
\begin{align}
R(t)&=N_n (1-t)^{\mu}(1+t)^{\nu} \;\; 2^{-2(\mu+\nu)}\;\; \frac{\Gamma(2\mu+1)\Gamma(n+1)}{\Gamma(2\mu+n+1)}  \notag \\
& \;\;\;\; \;\;P_n(2\mu,2\nu-1)(t).
\label{}
\end{align}
With regard to obtaining the normalisation constant $N_n$, we use the orthogonality relation of Jacobi polynomials (See Ref. \cite{Int1}, Eq. 7.391, page 806).
\begin{align}
N_n &=\bigg(\frac{2\tau\mu(\mu+\nu+n)}{\nu+n}\bigg)^\frac{1}{2}  \notag \\  & \;\;\;\;\;\;\bigg[\frac{n! \;\;\Gamma(2\mu+n+1) \; \Gamma(2\nu+2\mu+n)}     {(\Gamma(2\mu+1)\Gamma(n+1))^2 \; \Gamma(2\nu+n)}  \bigg]^\frac{1}{2}.
\label{}
\end{align}
\section{B(E2) Transition rates}
In general, the quadrupole operator is given by 
\begin{equation}
\begin{split}
T_{M}^{(E2)}= t\beta\bigg[ D_{M,0}^{(2)}(\theta_{i})\cos(\gamma-\frac{2\pi}{3})+\frac{1}{\sqrt2} \bigg(D_{M,2}^{(2)}(\theta_{i})   \;  \; \;    \\ + \; D_{M,-2}^{(2)}(\theta_{i}) \bigg) \sin(\gamma-\frac{2\pi}{3})         \bigg],
\end{split}
\label{}                                                                  
\end{equation}
where t is a scaler factor and $D_{M,\alpha}^{(2)}(\theta_{i})(\alpha=0,2,-2)$ denotes the Wigner functions of Euler angles. Around $\gamma=\frac{\pi}{6} $ (triaxial nuclei), this expression is simplified into
\begin{equation}
\begin{split}
T_{M}^{(E2)}= - \frac{1}{\sqrt2} \;t \beta \bigg(D_{M,2}^{(2)}(\theta_{i})+  D_{M,-2}^{(2)}(\theta_{i}) \bigg).    
\end{split}
\label{TB2}                                                                  
\end{equation}
The $B(E2)$ transition rates are given by \cite{Ed1}
\begin{equation}
\begin{split}
B(E2; L_{i}\alpha_{i} \rightarrow L_{f}\alpha_{f})=\frac{5}{16\pi}\frac{|<\L_{f}\alpha_{f}||T^{(E2)}||L_{i}\alpha_{i}>|^{2}}{(2L_{i}+1)},
\end{split}
\label{}                                                                  
\end{equation}
where the reduced matrix element is obtained through the Wigner-Eckart theorem \cite{Ed1}
\begin{align}
\begin{split}
|<\L_{f}\alpha_{f}|T_{M}^{(E2)}|L_{i}\alpha_{i}>|  &=\frac{(L_{i}2L_{f}|\alpha_{i}M\alpha_{f})}{\sqrt{2L_{f}+1}}  \label{BME}                                                                  
\\  
& \;\;\; \times|<\L_{f}\alpha_{f}||T^{(E2)}||L_{i}\alpha_{i}>|.
\end{split}
\end{align}
In equation \eqref{BME}, the integral over $\gamma$ leads to unity (because of the normalisation of $\Gamma(\gamma)$ ), the integral over the Euler Angles is performed by using the standard integrals of three Wigner functions and the integral over $\beta$ takes the form
\begingroup\small
\begin{align}
I_{\beta}(n_{i},L_{i},\alpha_{i},n_{f},L_{f},\alpha_{f}) &=  \\
& \int_0^\infty \beta \; \xi_{n_i,L_i,\alpha_i}(\beta) \; \xi_{n_f,L_f,\alpha_f}(\beta)\; \beta^4 \; d\beta, \nonumber
\label{}                                                                  
\end{align}
\endgroup
where the factor $\beta^4$  comes from the volume element \cite{B1} and the factor $\beta$  comes from Eq.\eqref{TB2}. The final result gives the general expression of $E2$ transition probabilities

\begin{align}
B(E2; L_{i}\alpha_{i} \rightarrow L_{f}\alpha_{f}) &= \frac{5}{16\pi}\frac{t^{2}}{2} 
 \frac{1}{(1+\delta_{\alpha_i,0})(1+\delta_{\alpha_f,0})} \notag \\
&  \bigg[   (L_i2L_f|\alpha_i2\alpha_f)+ (L_i2L_f|\alpha_i-2\alpha_f) \notag  \\
& + (-1)^{L_i}(L_i2L_f|-\alpha_i2\alpha_f)   \bigg]^{2}  \notag \\
& I^{2}_{\beta}(n_{i},L_{i},\alpha_{i},n_{f},L_{f},\alpha_{f}).  
\label{CGs}                                                                  
\end{align}
The Clebsch-Gordan coefficients (CGCs) appearing in equation \eqref{CGs} impose a $\Delta\alpha=\pm2$ selection rule. We use the derived ratios given in Ref. \cite{Bo2} in order to calculate the transitions rates \\
\hspace*{0.3 cm} a) within the g.s band, \\
\hspace*{0.3 cm} b) within the even levels of the $\gamma$ band, \\
\hspace*{0.3 cm} c) within the odd levels of the $\gamma$ band, \\
\hspace*{0.3 cm} d) between the even levels of $\gamma$ band and the g.s band, \\
\hspace*{0.3 cm} e) between the odd levels of $\gamma$ band and the g.s band,   \\
\hspace*{0.3 cm} f) between the odd levels  and the even levels of $\gamma$ band.
\section{Numerical results}
In this section, we present all results we have obtained with  Z(5)-HD model. This model was applied for calculating  energy ratios of excited collective states and  reduced $E2$ transitions probabilities for the isotopes  $^{126,128,130,132,134}$Xe and $^{192,194,196}$Pt. Such nuclei have been previously chosen to be studied within other models  \cite{BO4,Ch2,BO6} because of  undergoing the signature of the triaxial rigid rotor \cite{Ra,Da}
\begin{equation}
\Delta E=|E_{{2}^{+}_{g}}+E_{{2}^{+}_{\gamma}}-E_{{3}^{+}_{\gamma}}| \approx 0.
\label{De1}                                                                  
\end{equation}
This equation is used in an approximate way, because the experimental data for the eight nuclei $^{126,128,130,132,134}$Xe and $^{192,194,196}$Pt, respectively lead to the values
\begin{equation}
\Delta E(keV)=49, 17, 26, 162, 379, 8, 28, 29.
\label{De2}                                                                  
\end{equation}
 By referring to the values of equation \eqref{De2}, the isotopes $^{128,130}$Xe and $^{192,194,196}$Pt  are good candidates for a triaxial rotor model and hence, presumptively could present triaxial deformation in their structure. Such a feature will be checked afterwards through another important signature. So, the above formula \eqref{De1} served us in the first step as a guide in choosing the candidate nuclei and therefore, we have added the isotopes $^{126,132,134}$Xe  in our present study. \\
The allowed bands (i.e. ground state (g.s), $\beta$ and $\gamma$) are labelled by the quantum numbers, $n$, $n_{w}$, $n_{\gamma}$ and $L$. As described in the framework of the rotation-vibration model \cite{Gr}, the lowest bands for $Z(5)$ are as follows 
\begin{enumerate}
\item The g.s band  is characterized by $n = 0$, $n_{\gamma}=0$, $n_{w}=0$
\item The $\beta$ band is characterized by $n = 1$, $n_{\gamma}=0$, $n_{w}=0$.
\item The $\gamma$ band composed by the even $L$ levels with $n =0$, $n_{\gamma}=0$, $n_{w}=2$ and the odd $L$ levels with $n = 0$, $n_{\gamma}=0$, $n_{w}=1$.
\end{enumerate}
Discussion on nature of such bands can be found in the recent review article \cite{Shar19}. \\
The energy spectrum is given by equation \eqref{En} and depends on four parameters, namely: the screening parameter $\tau$ in the $\beta$ potential,  the ring-shape parameters $c$ and $s$ of the $\gamma$ potential and the mass deformation parameter $a$. Our task is to fit these parameters to reproduce the experimental data by applying a least-squares fitting procedure for each considered isotope. We evaluate the root mean square (r.m.s) deviation between the theoretical values and the experimental data by
\begin{equation}
\sigma=\sqrt{\frac{\sum_{i=1}^{m}(E_{i}(exp)-E_{i}(th))^2}{(m-1)E(2_{1}^{+})^2}},
\label{De3}                                                                  
\end{equation}
where $E_{i}(exp)$ and $E_{i}(th)$ represent the experimental and theoretical energies of the $\textit i^{th}$ level, respectively, while $m$ denotes the number of states. $E(2_{1}^{+})$ is the energy of the first excited level of the g.s band.
The corresponding free parameters ($\tau$, $c$, $s$) and the mass deformation parameter $a$ are listed in table \ref{table:T1}. 
In this table, we give the fitted parameters allowing to reproduce the experimental data \cite{ww1} and $Z(5)$ model \cite{Bo2}. The results presented here have been obtained for $\delta=\lambda=0$. Different choices for $\delta$ and $\lambda$   lead to a renormalization of the parameters values $\tau$, $c$, $s$ and $a$ , so that the predicted energy levels remain unchanged.
\begin{table}[h]
 	\small\noindent\tabcolsep=6pt
 	\begin{tabular}{ c c c c c c c c c}
 		\hline 
 		
 		\hline
 		\\[-8pt]
 		nuclei \qquad&${\tau}$& $c$& $s$& $a$& $L_{g}$& $L_{\beta}$& $L_{\gamma}$& $m$ \\
 		\hline
 		\\[-8pt]
 		{$^{126}$Xe} \quad&0.071& 8& 192& 0.0025& 12& 4& 9& 16 \\
 		{$^{128}$Xe} \quad&0.050& 2& 140& 0.0000& 10& 2& 7& 12 \\
 		{$^{130}$Xe} \quad&0.010& 0& 140& 0.0000& 14& 0& 5& 11 \\
 		{$^{132}$Xe} \quad&0.080& 72& 226& 0.0000& 6&  0& 5& 7 \\
 		{$^{134}$Xe} \quad&0.080& 78& 187& 0.0000& 6&  0& 5& 7 \\
 		{$^{192}$Pt} \quad&0.050& 19& 73& 0.0010& 10& 4& 8& 14 \\
 		{$^{194}$Pt} \quad&0.059& 6& 195& 0.0030& 10& 4& 8& 13 \\
 		{$^{196}$Pt} \quad&0.086& 7& 120& 0.0059& 10& 4& 8& 13 \\
 		{$Z(5)$}     \quad&0.039& 11& 406& -& 14& 4& 9& 17 \\
 		\hline
 	\end{tabular}
 	\caption{The Hulth\'{e}n potential and deformation parameters values fitted to the experimental data \cite{ww1} as well as the results of $Z(5)$ model \cite{Bo2}.  $L_{g}$, $L_{\beta}$ and $L_{\gamma}$ characterize the angular momenta of the highest levels of the ground state, $\beta$ and $\gamma$ bands respectively, included in the fit, while $m$ is the total number of experimental states involved in the r.m.s fit.}
\label{table:T1}
 \end{table}
\\
In Ref \cite{BO4} (Z(5)-DD model), the authors have presented the analytical results for triaxial nuclei with $\gamma=\frac{\pi}{6}$ by using Davidson potential, but they did not present their numerical results. So, in order to compare our results obtained with Hulth\'{e}n potential with those obtained with Davidson potential, we have used the analytical formulas of Ref \cite{BO4} and all obtained numerical results  are presented in table  \ref{table:T2}. Here, the parameters $\beta_{0}$, $c$ and $a$  are respectively the Davidson potential parameter, the $\gamma$-potential parameter  and the  mass deformation parameter.

\begin{table}[h]
 	\small\noindent\tabcolsep=6pt
 	\begin{tabular}{ c c c c  c c c c}
 		\hline 
 		
 		\hline
 		\\[-8pt]
 		nuclei \qquad&${\beta_{0}}$& $c$& $a$& $L_{g}$& $L_{\beta}$& $L_{\gamma}$& $m$ \\
 		\hline
 		\\[-8pt]
 		{$^{126}$Xe} \quad&1.19& 1.38&  0.0060& 12& 4& 9& 16 \\
 		{$^{128}$Xe} \quad&0.94& 4.52&  0.0000& 10& 2& 7& 12 \\
 		{$^{130}$Xe} \quad&0.11& 4.53&  0.0000& 14& 0& 5& 11 \\
 		{$^{132}$Xe} \quad&0.00& 0.00&  0.0000& 6&  0& 5& 7 \\
 		{$^{134}$Xe} \quad&0.00& 0.00&  0.0000& 6&  0& 5& 7 \\
 		{$^{192}$Pt} \quad&1.05& 8.47&  0.0035& 10& 4& 8& 14 \\
 		{$^{194}$Pt} \quad&1.04& 6.88&  0.0055& 10& 4& 8& 13 \\
 		{$^{196}$Pt} \quad&0.84& 3.05&  0.0093& 10& 4& 8& 13 \\
 		{$Z(5)$}     \quad&1.37& 7.52&  -& 14& 4& 9& 17 \\
 		\hline
 	\end{tabular}
 	\caption{The Davidson potential and deformation parameters values fitted to the experimental data \cite{ww1} as well as the results of $Z(5)$ model \cite{Bo2}. $L_{g}$, $L_{\beta}$ and $L_{\gamma}$ characterize the angular momenta of the highest levels of the ground state, $\beta$ and $\gamma$ bands respectively, included in the fit, while $m$ is the total number of experimental states involved in the rms fit.}
\label{table:T2}
 \end{table}
  
From table \ref{table:T1}, containing the results obtained with  Hulth\'{e}n potential,  the following remarks are applying: \\
\begin{enumerate}
\item For the isotope $^{126}$Xe, the term of the deformation becomes necessary, leading to nonzero value from the fit. \\
\item  For the isotopes $^{128,130,132,134}$Xe, however, the fitting  leads  to a mass deformation parameter $a=0$. So, they are vibrational isotopes. \\
\item  For the isotopes $^{192,194,196}$Pt, the fitting leads to non zero values of the mass deformation parameter which increase with mass number $A$.
\end{enumerate} 
From table  \ref{table:T2}, the results are obtained with Davidson potential and the following remarks are applying:
\begin{enumerate}
\item For $^{126}$Xe, the $\beta_{0}$ and $a$ terms become necessary, leading to non zero values of both them.
\item  For $^{128,130}$Xe, the fitting leads to non zero $\beta_{0}$, however the mass deformation parameter  $a=0$. 
\item  For $^{132,134}$Xe, both $\beta_{0}$ and $a$ parameters are equal to zero. Therefore, there is no need for deformation dependence in the potential. 
\item  For $^{192,194,196}$Pt, the fitting leads to non zero values of  $\beta_{0}$ and $a$, and the mass deformation parameter increases with mass number $A$.
\end{enumerate}
 
\begin{figure}[h]
      	\begin{center}
      		\includegraphics[scale=0.7]{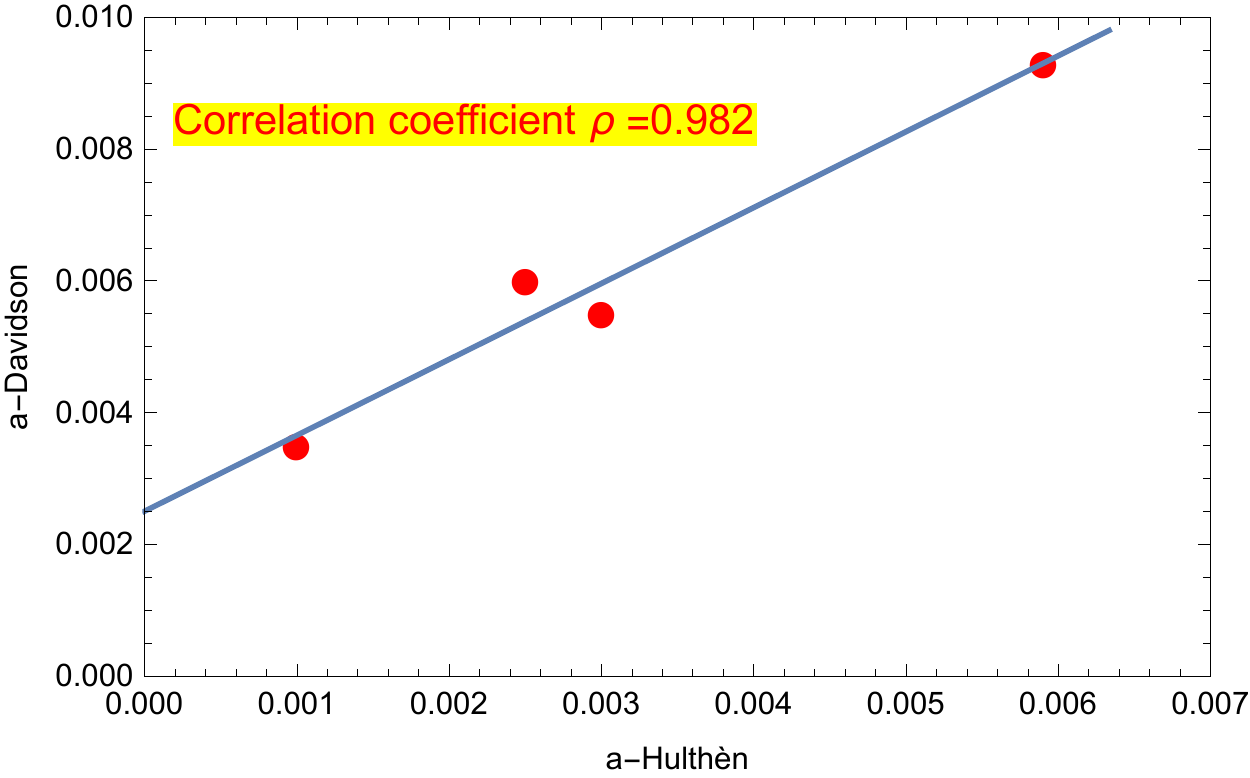}
      		\caption{The comparison of  deformation parameter $a$ given by our model Z(5)-HD and Z(5)-DD model \cite{BO4} for isotopes $^{126}$Xe and $^{192,194,196}$Pt.}
      		\label{fig0}
      	\end{center}
      \end{figure}

\begin{figure}
      	\begin{center}
      		\includegraphics[scale=0.9]{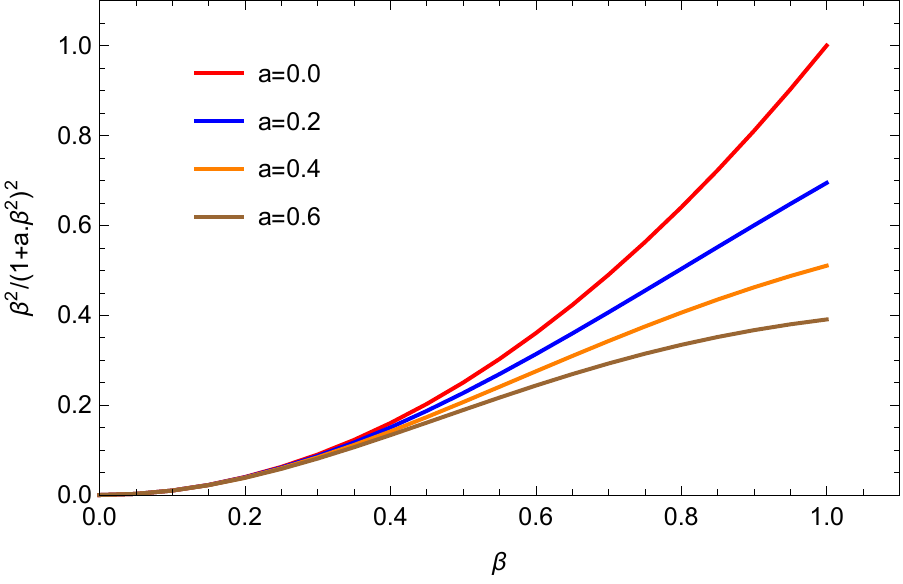}
      		\caption{The function $\beta^2/f^2(\beta)={\beta^2}/{(1+a\beta^2)^2} $ plotted as a function of the nuclear deformation $\beta$ for different arbitrary values of  parameter $a$}
      		\label{fig100}
      	\end{center}
      \end{figure}

\begin{figure}
      	\begin{center}
      		\includegraphics[scale=0.9]{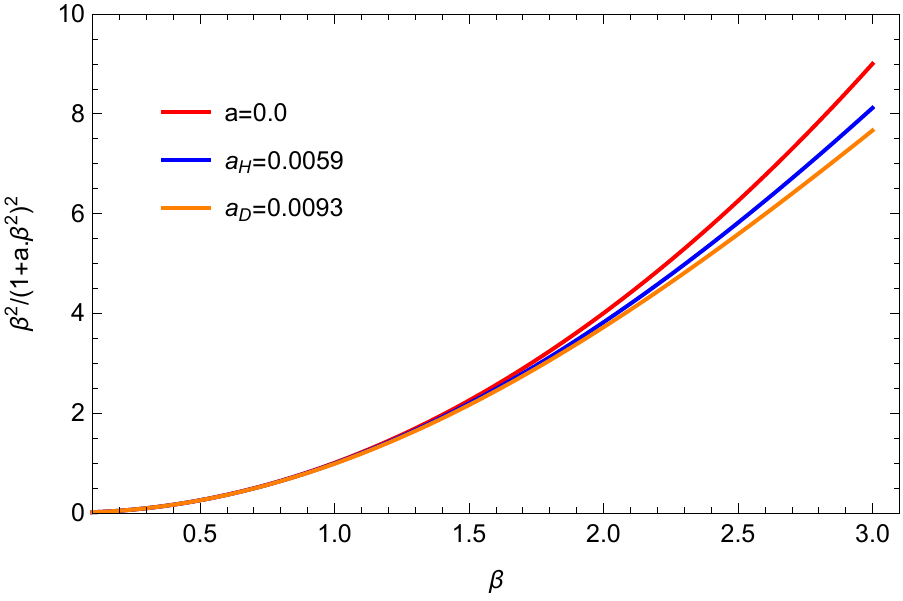}
      		\caption{The function $\beta^2/f^2(\beta)={\beta^2}/{(1+a\beta^2)^2} $ plotted as a function of the nuclear deformation $\beta$ for values of the parameter $a$ obtained for $^{196}$Pt isotope with Z(5)-HD and Z(5)-DD models }
      		\label{fig101}
      	\end{center}
      \end{figure}

\begin{figure*}
      	\begin{center}
      		\includegraphics[scale=0.228]{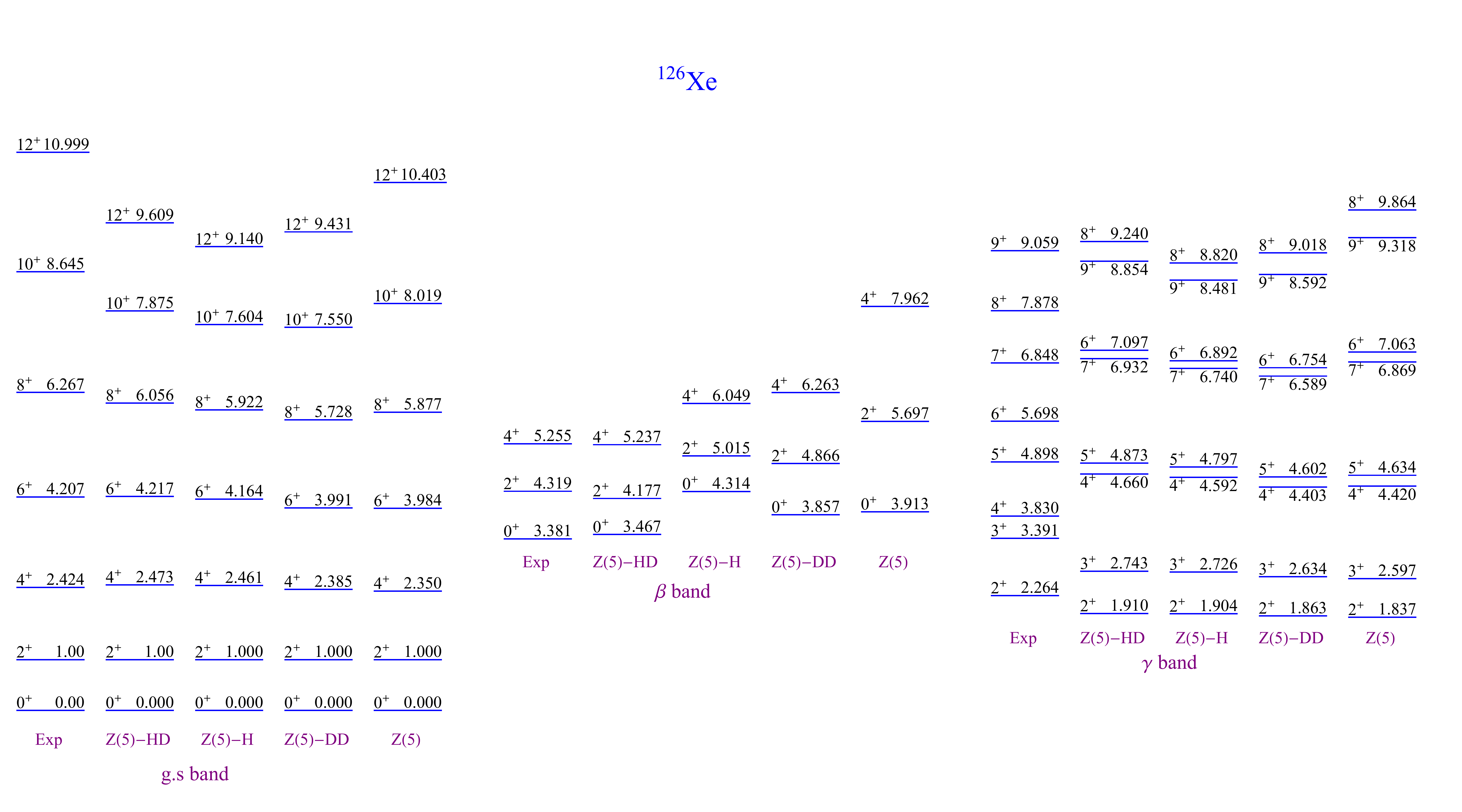}
      		\caption{The comparison between our theoretical energy spectra, given by Eq. \eqref{En} using the parameters in Table \ref{table:T1} for $^{126}$Xe isotope with the experimental data \cite{ww1}, those obtained in Ref. \cite{Ch2} and in Ref. \cite{BO4} using parameters in table \ref{table:T2} and those from free parameters model Z(5)\; \cite{Bo2}.}
      		\label{fig1}
      	\end{center}
      \end{figure*}

\begin{figure*}
      	\begin{center}
      		\includegraphics[scale=0.228]{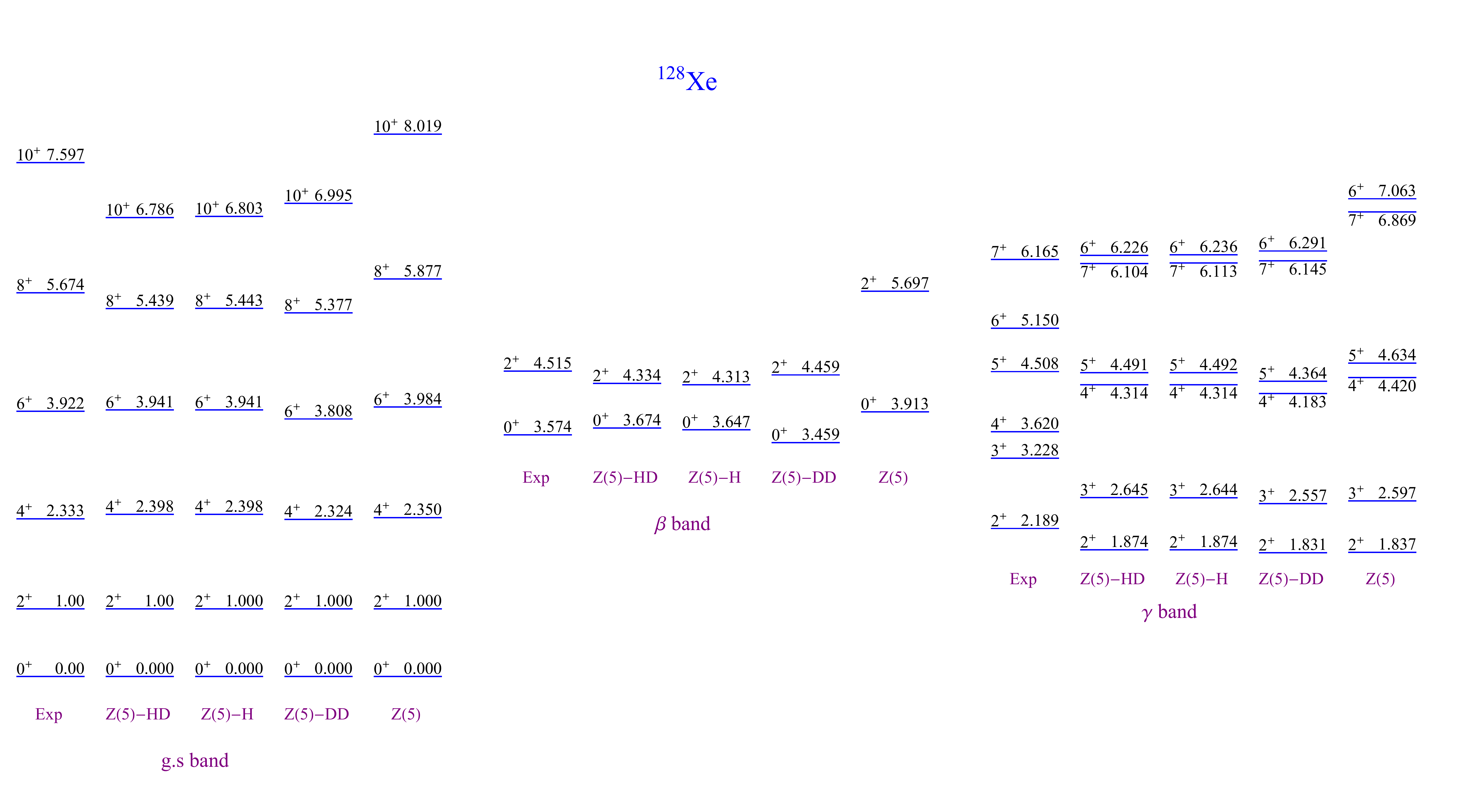}
      		\caption{The comparison between our theoretical energy spectra, given by Eq. \eqref{En} using the parameters in Table \ref{table:T1} for $^{128}$Xe isotope with the experimental data \cite{ww1}, those obtained in Ref. \cite{Ch2} and in Ref. \cite{BO4} using parameters in table \ref{table:T2} and those from free parameters model Z(5)\; \cite{Bo2}.}
      		\label{fig2}
      	\end{center}
      \end{figure*}

\begin{figure*}
      	\begin{center}
      		\includegraphics[scale=0.228]{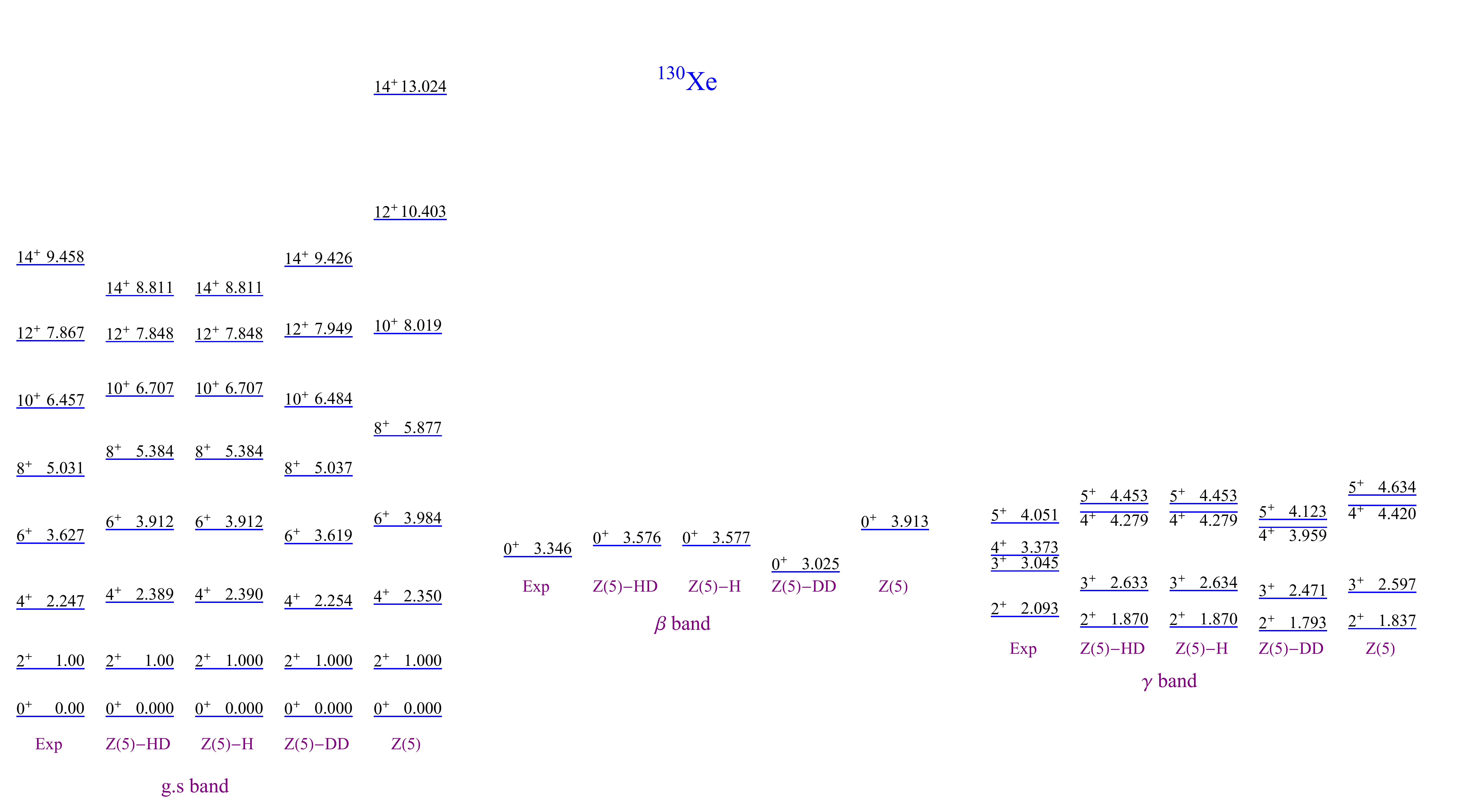}
      		\caption{The comparison between our theoretical energy spectra, given by Eq. \eqref{En} using the parameters in Table \ref{table:T1} for $^{130}$Xe isotope with the experimental data \cite{ww1}, those obtained in Ref. \cite{Ch2} and in Ref. \cite{BO4} using parameters in table \ref{table:T2} and those from free parameters model Z(5)\; \cite{Bo2}.}
      		\label{fig3}
      	\end{center}
      \end{figure*}

\begin{figure*}
      	\begin{center}
      		\includegraphics[scale=0.228]{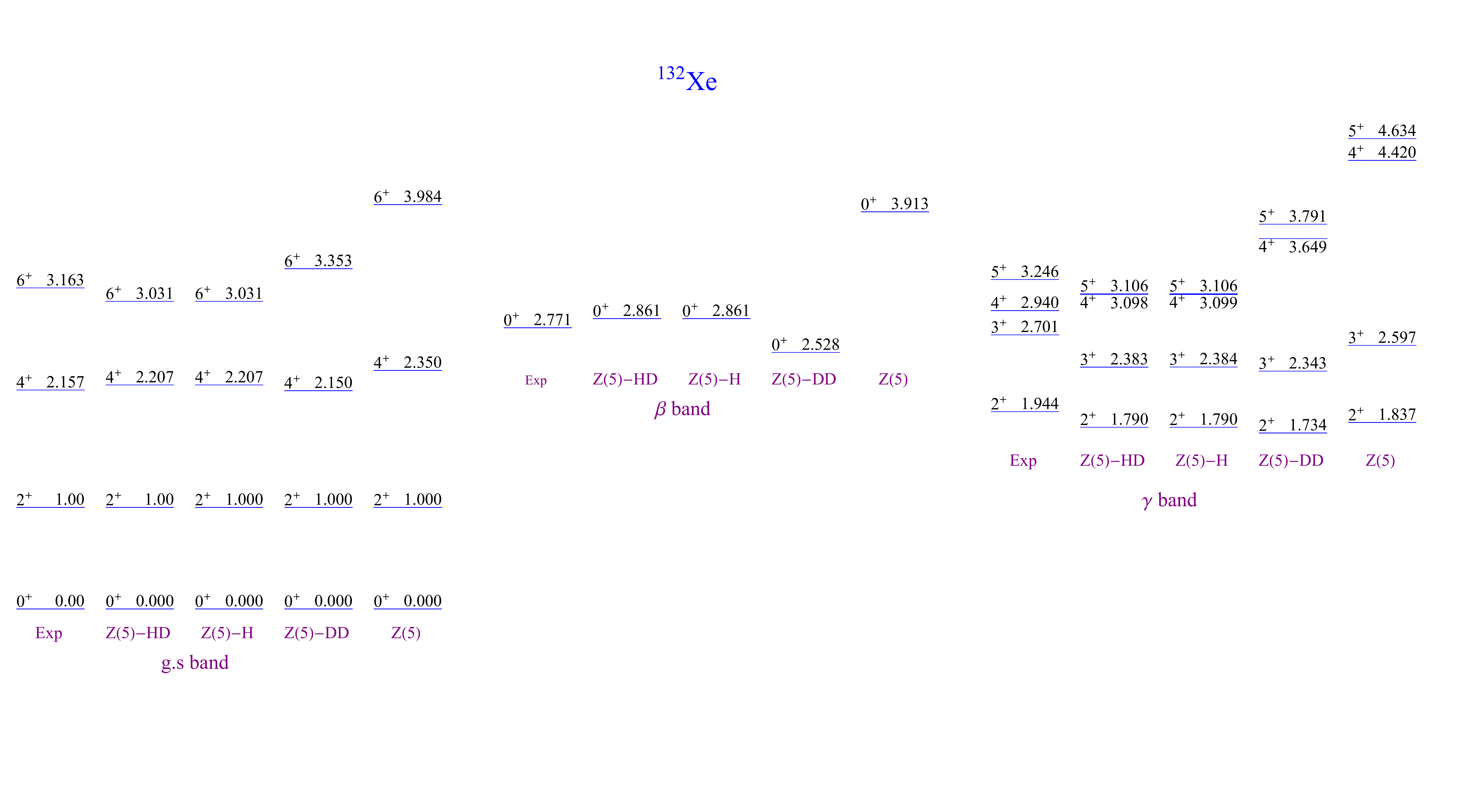}
      		\caption{The comparison between our theoretical energy spectra, given by Eq. \eqref{En} using the parameters in Table \ref{table:T1} for $^{132}$Xe isotope with the experimental data \cite{ww1}, those obtained in Ref. \cite{Ch2} and in Ref. \cite{BO4} using parameters in table \ref{table:T2} and those from free parameters model Z(5)\; \cite{Bo2}.}
      		\label{fig4}
      	\end{center}
      \end{figure*}

\begin{figure*}
      	\begin{center}
      		\includegraphics[scale=0.228]{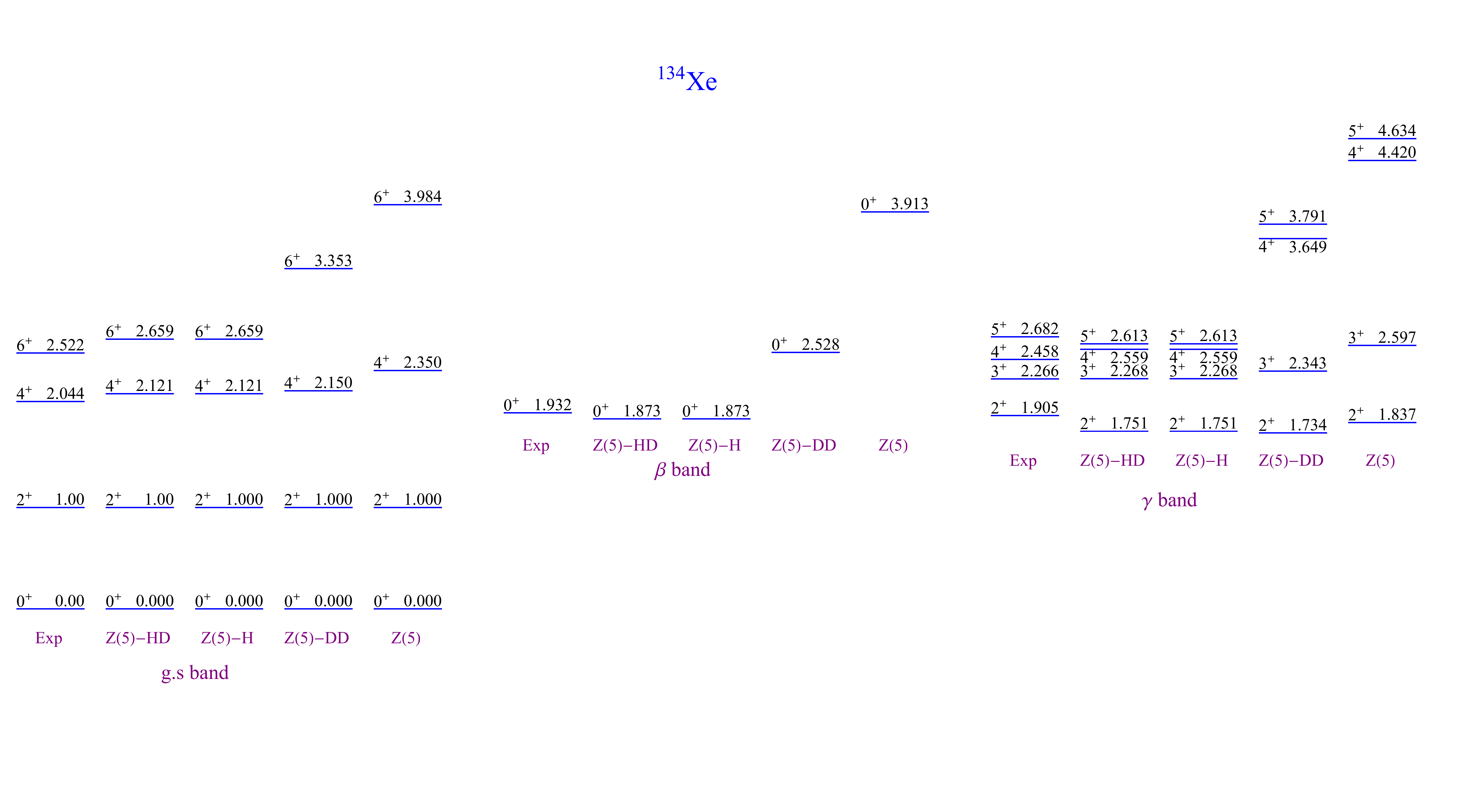}
      		\caption{The comparison between our theoretical energy spectra, given by Eq. \eqref{En} using the parameters in Table \ref{table:T1} for $^{134}$Xe isotope with the experimental data \cite{ww1}, those obtained in Ref. \cite{Ch2} and in Ref. \cite{BO4} using parameters in table \ref{table:T2} and those from free parameters model Z(5)\; \cite{Bo2}.}
      		\label{fig5}
      	\end{center}
      \end{figure*}

\begin{figure*}
      	\begin{center}
      		\includegraphics[scale=0.228]{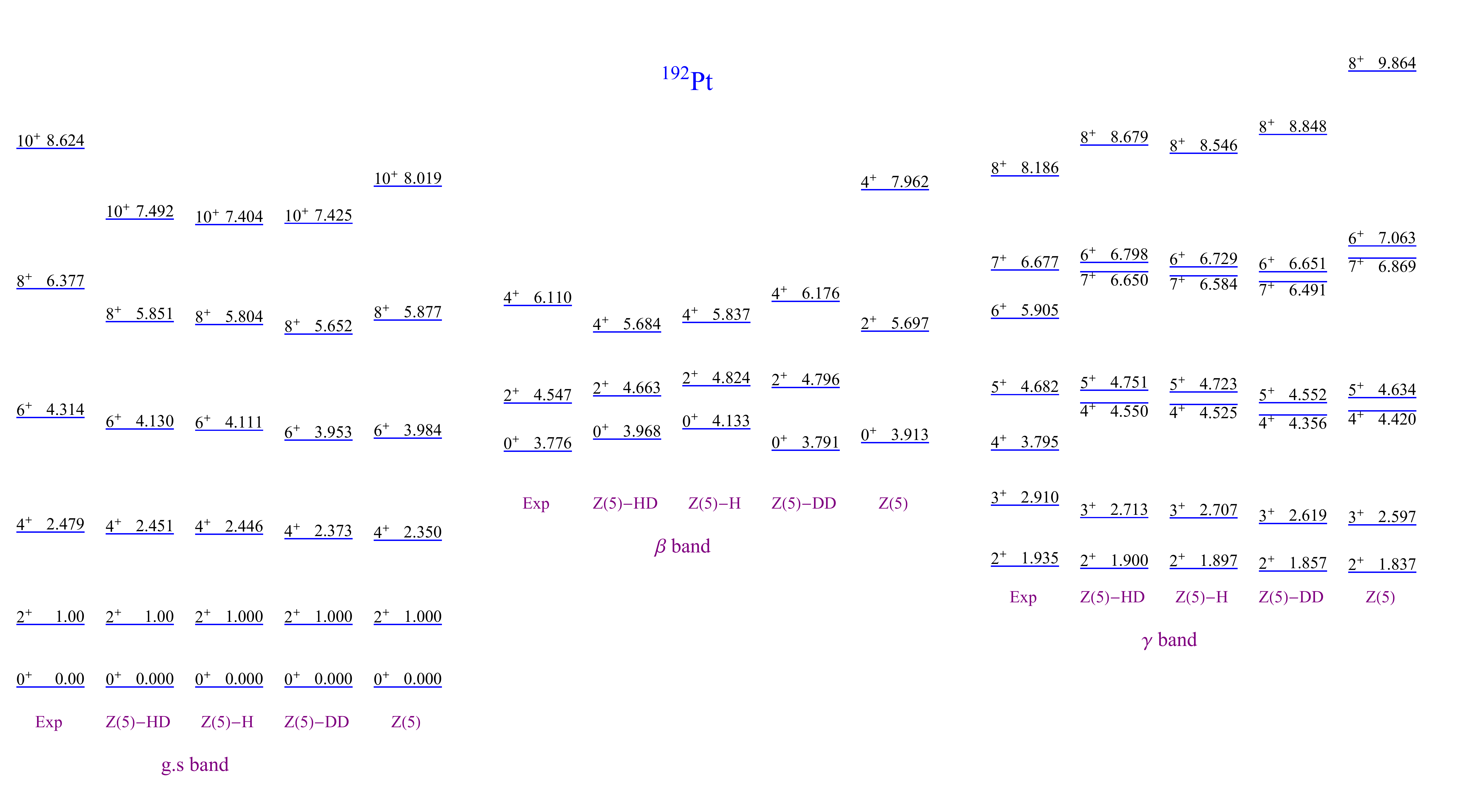}
      		\caption{The comparison between our theoretical energy spectra, given by Eq. \eqref{En} using the parameters in Table \ref{table:T1} for $^{192}$Pt isotope with the experimental data \cite{ww1}, those obtained in Ref. \cite{Ch2} and in Ref. \cite{BO4} using parameters in table \ref{table:T2} and those from free parameters model Z(5)\; \cite{Bo2}.}
      		\label{fig6}
      	\end{center}
      \end{figure*}

\begin{figure*}
      	\begin{center}
      		\includegraphics[scale=0.228]{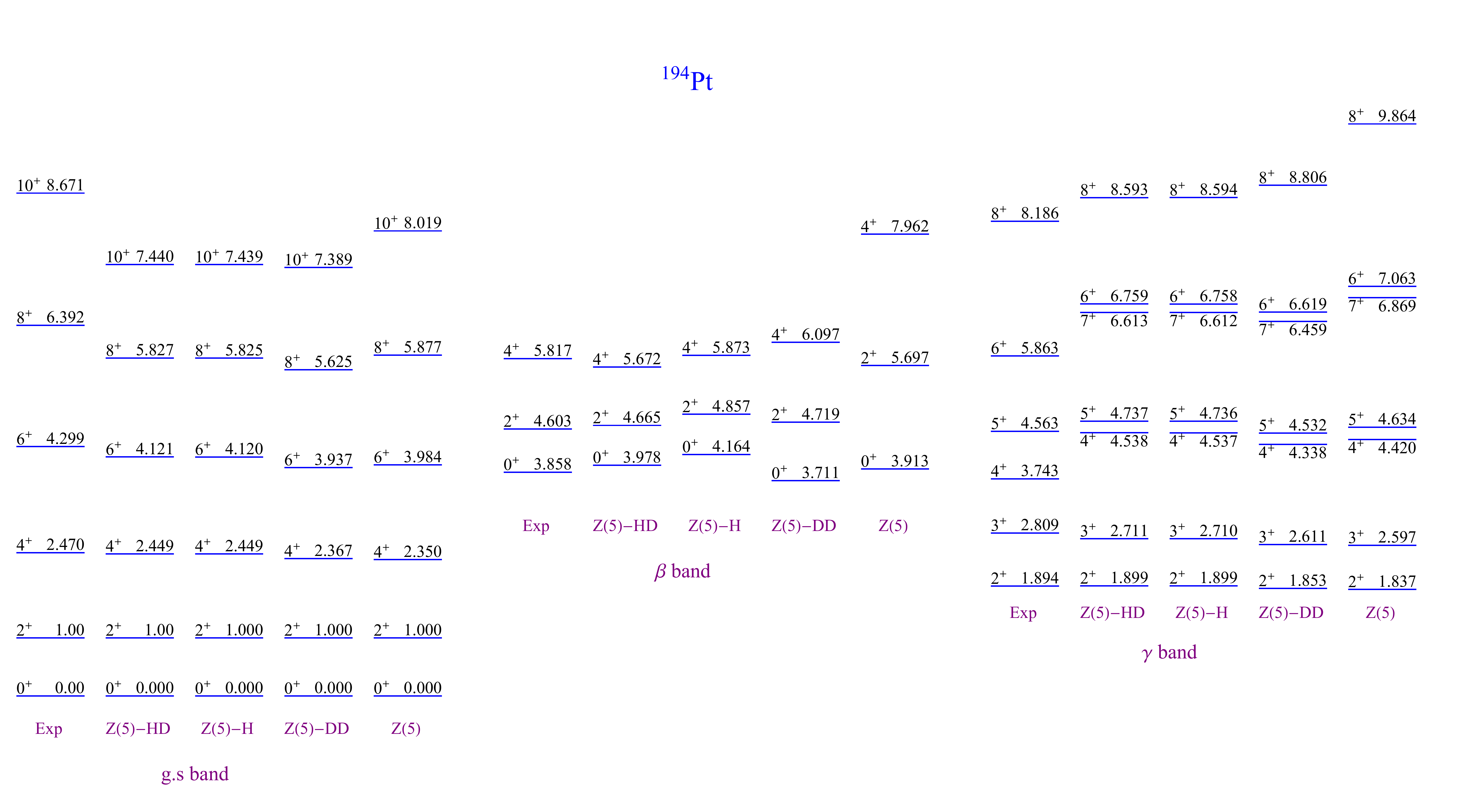}
      		\caption{The comparison between our theoretical energy spectra, given by Eq. \eqref{En} using the parameters in Table \ref{table:T1} for $^{194}$Pt isotope with the experimental data \cite{ww1}, those obtained in Ref. \cite{Ch2} and in Ref. \cite{BO4} using parameters in table \ref{table:T2} and those from free parameters model Z(5)\; \cite{Bo2}.}
      		\label{fig7}
      	\end{center}
      \end{figure*}

\begin{figure*}
      	\begin{center}
      		\includegraphics[scale=0.228]{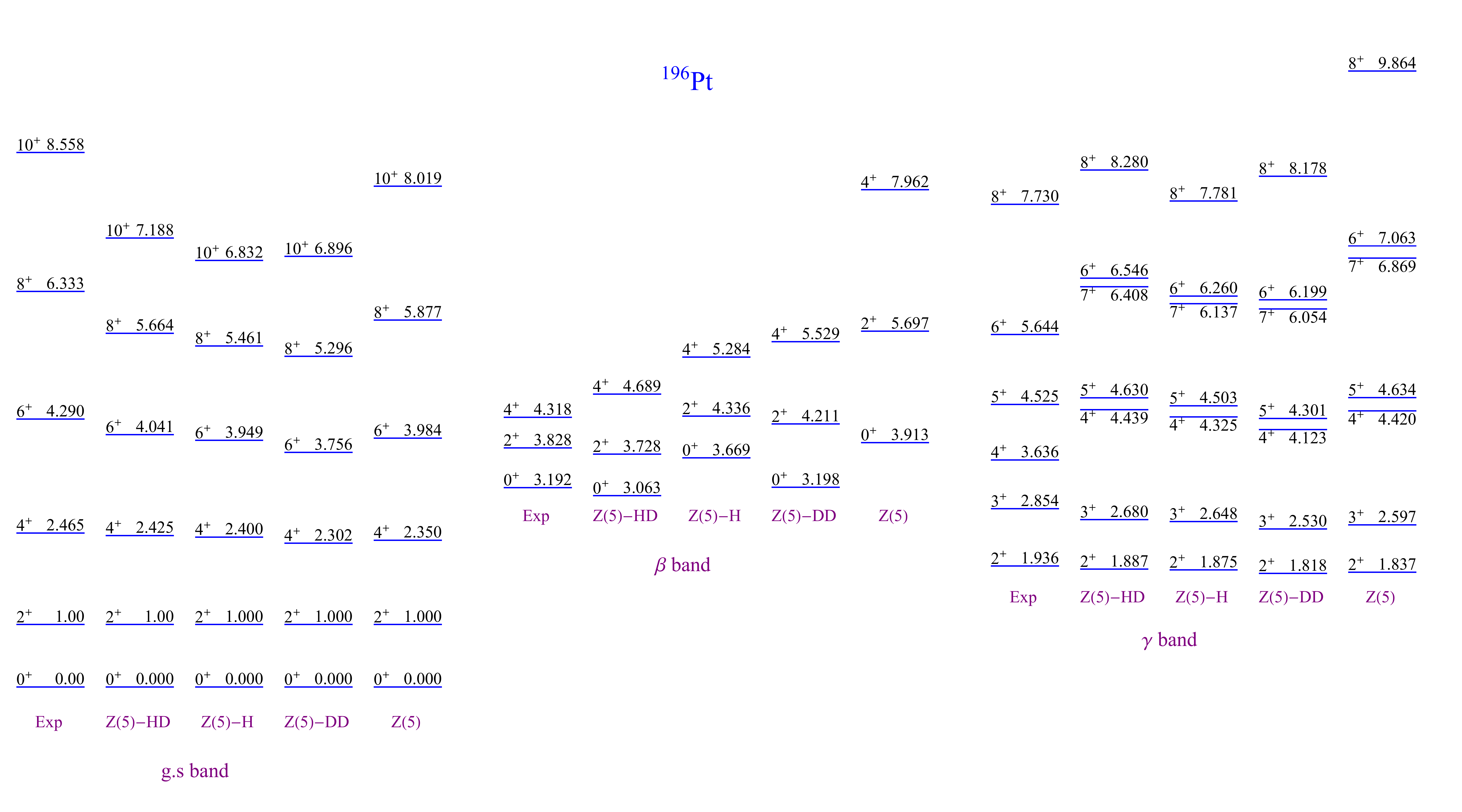}
      		\caption{The comparison between our theoretical energy spectra, given by Eq. \eqref{En} using the parameters in Table \ref{table:T1} for $^{196}$Pt isotope with the experimental data \cite{ww1}, those obtained in Ref. \cite{Ch2} and in Ref. \cite{BO4} using parameters in table \ref{table:T2} and those from free parameters model Z(5)\; \cite{Bo2}.}
      		\label{fig8}
      	\end{center}
      \end{figure*}

From results displayed in tables \ref{table:T1} and \ref{table:T2}, we conclude that $^{128,130}$Xe are vibrational isotopes and $^{132,134}$Xe are pure vibrators (both parameters $\beta_{0}$ and $a$ are null). Thus, our results confirm those obtained in Refs \cite{BO4,BO6} concerning the vibrational nature of these isotopes. Such a result is corroborated by the obtained values below for the ratio $R_{4/2}$ which are close to the vibrator's characteristic value: 2. \\
In Fig.\ref{fig0}, we compare the values of the mass deformation parameter given by our model Z(5)-HD and those obtained with Z(5)-DD \cite{BO4} model for $^{126}$Xe and $^{192,194,196}$Pt. We note that there is a strong correlation between them ($\rho=0.982$). So, as it has been previously mentioned in Ref.\cite{Ch5}, the mass deformation parameter $a$ is not a simple fitting parameter to be adjusted, but a structural parameter of the model. It is not influenced by the presence of the parameters of the used potential. Moreover, this parameter plays an important role for the moment of inertia inasmuch as it moderates the variation of the latter when the nuclear deformation $\beta$ increases as can be seen from Fig.\ref{fig100} with arbitrary values of $a$ and Fig.\ref{fig101} for a concrete case.\\   
  In table \ref{table:T3}, we compare the quality measure $\sigma$ of our results (Z(5)-HD) with that of Z(5)-H model \cite{Ch2}, Z(5)-DD model  \cite{BO4} and Z(5) model \cite{Bo2}. 
\begin{table}[h]	
 	\small\noindent\tabcolsep=6pt
 	\begin{tabular}{ c c c c c }
 		\hline 
 		
 		\hline
 		\\[-8pt]
 		nuclei \qquad&Z(5)-HD& Z(5)-H& Z(5)-DD& Z(5) \\
 		\hline
 		\\[-8pt]
 		{$^{126}$Xe} \quad&0.716& 0.835& 0.791&  1.082\\
 		{$^{128}$Xe} \quad&0.508& 0.508& 0.495&  0.802\\
 		{$^{130}$Xe} \quad&0.443& 0.443& 0.297&  1.564\\
 		{$^{132}$Xe} \quad&0.181& 0.181& 0.422&  1.013\\
 		{$^{134}$Xe} \quad&0.123& 0.123& 0.790&  1.524\\
 		{$^{192}$Pt} \quad&0.517& 0.521& 0.528&  0.886 \\
 		{$^{194}$Pt} \quad&0.544& 0.553& 0.566&  0.973\\
 		{$^{196}$Pt} \quad&0.602& 0.718& 0.746&  1.448\\
 		\hline
 	\end{tabular}
 	\caption{The root mean square (rms) deviation  between experimental data \cite{ww1} and the theoretical results corresponding to Z(5)-HD, Z(5)-H \cite{Ch2}, Z(5)-DD  \cite{BO4} and Z(5) \cite{Bo2}  of given isotopes.}
\label{table:T3}
 \end{table}
From this table, one can see that our model is generally more efficient than the all others. Moreover, one can observe that the r.m.s for the two isotopes $^{132}$Xe and $^{134}$Xe are equal for both models Z(5)-HD and Z(5)-H and so smaller than that for Z(5)-DD. Hence, we can conclude that  Hulth\'{e}n potential is more suitable for describing pure vibrators than the Davidson one.\\
In figures \ref{fig1}-\ref{fig8}, we have plotted the energy spectra of the isotopes $^{126,128,130,32,134}$Xe and $^{192,194,196}$Pt. From Fig.\ref{fig1}, one can see that in the g.s band of $^{126}$Xe, our model Z(5)-HD reproduces well the experimental levels in comparison with Z(5)-H and Z(5)-DD. Also, in the $\beta$ band, Z(5)-HD is more precise than the others, while in the $\gamma$ band the difference between all models' calculations is not significant.\\
Fig.\ref{fig2} shows the spectrum of $^{128}$Xe, where our model is still more efficient for mostly all levels  in the g.s band except the last one, while in $\beta$ and $\gamma$ bands there is generally no significant difference between all models. However, one can observe that the $\beta$ band head is better reproduced with both Z(5)-HD and Z(5)-H models.\\
The spectrum of $^{130}$Xe presented in Fig.\ref{fig3} shows that the levels from $L=2$ to $L=4$, in the g.s band, are well reproduced by all models. Nevertheless, for the levels above $L=4$ except the one with $L=12$, Z(5)-DD is more precise than the others. However, the $\beta$ band head is better reproduced with Z(5)-HD and Z(5)-H than Z(5)-DD. As to the $\gamma$ band, all models are almost equal in the reproduction of all levels except the levels $L=4$ and $L=5$ where Z(5)-DD is slightly more precise.\\
From Fig.\ref{fig4} and Fig.\ref{fig5}, representing the energy spectra for $^{132}$Xe and $^{134}$Xe, one can see that the calculations of Z(5)-HD and Z(5)-H are identical because in this case, the mass deformation parameter is null. Besides, these results are fairly better than those obtained within Z(5)-DD. So, as it was already mentioned above, the Hulth\'{e}n potential is more appropriate for nuclei possessing a vibrational nature than the Davidson potential.\\   
Fig.\ref{fig6} presents the spectrum of $^{192}$Pt. Here, one can see that, in the g.s band, all levels are well reproduced by all models, but with some prevalence of Z(5)-HD followed by Z(5)-H except the last level. In the $\beta$ band, the levels $0^{+}$ and $4^{+}$ are better calculated with Z(5)-DD, but the $2^{+}$ is well reproduced with Z(5)-HD. In the $\gamma$ band, our model Z(5)-HD followed by Z(5)-H show some performance in respect to Z(5)-DD except for levels $4^{+}$ and $6^{+}$.\\
From the spectra of $^{194}$Pt and $^{196}$Pt given respectively in Fig.\ref{fig7} and Fig.\ref{fig8}, we can make the same observation in the g.s band like for the isotope $^{192}$Pt. However, in the $\beta$ band of $^{194}$Pt, all levels are well described with Z(5)-HD in respect to the others, while for $^{196}$Pt, the calculations of Z(5)-HD are the most precise followed by those of Z(5)-DD. As to the $\gamma$ band, the situation for both isotopes is similar to that of $^{192}$Pt. But, here, we have to notice that the common feature of all presented spectra is the observed inversion of the levels $6^{+}$ and $7^{+}$. The origin of this effect has been already explained in  Ref. \cite{Ch2}   where it has been also mentioned that such a feature appears just in spectra of triaxial nuclei and hence could be considered as a signature of triaxiality in nuclei. \\

\begin{figure*}
      	\begin{center}
      		\includegraphics[scale=1.228]{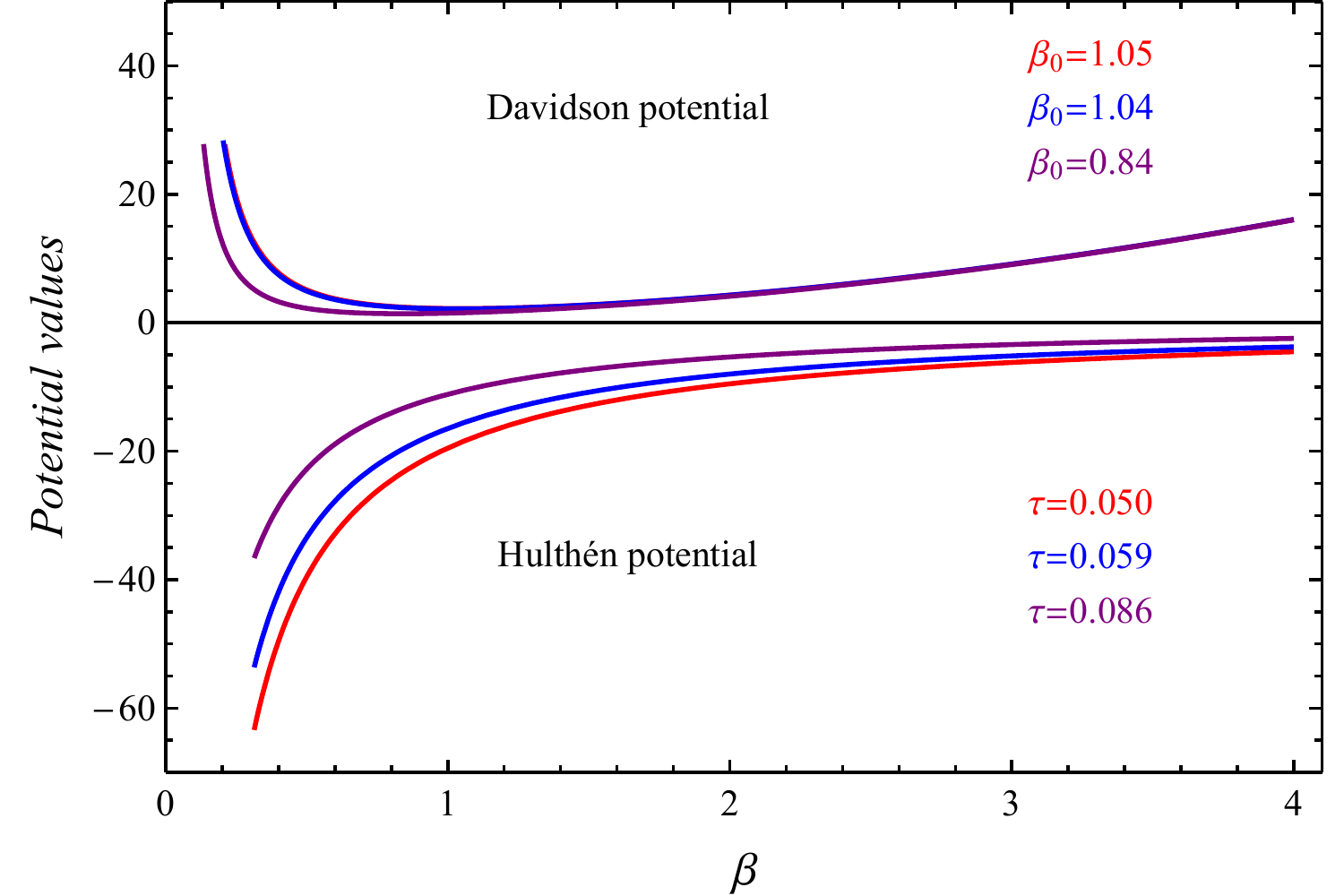}
      		\caption{The Hulthén and Davidson potentials with parameters given in tables \ref{table:T1} and \ref{table:T2} for $^{192,194,196}$Pt  isotopes }
      		\label{fig9}
      	\end{center}
      \end{figure*}   
By using the potential and deformation parameters values ( $\tau$, $c$, $s$, $a$) given in table \ref{table:T1} for Hulth\'{e}n potential and ($\beta_{0}$, $c$, $a$) given in table \ref{table:T2} for Davidson potential, which are obtained by fitting the energy ratios, we have calculated the intra-band and inter-band $B(E2)$ transition rates, normalized to  $B(E2;2_{0,0}^{+} \rightarrow 0_{0,0}^{+})$. Let us simply note that the reduced $E2$ transition probabilities have not been taken into account in the fitting procedure. However, to calculate the $B(E2)$ transition rates for the isotopes $^{128,130,132,134}$Xe, in the case of Davidson potential, exceptionally we have used the analytical outcome given in Ref \cite{BO6}. All our results are presented in tables \ref{table:T4}  and \ref{table:T5} alongside with those obtained with Z(5)-H, Z(5)-DD, esM and Z(5) models as well as the experimental data. From these tables, one can make the following observations:

\hspace{-0.2 cm} 1) For transitions between the lower levels, our theoretical results obtained with Z(5)-HD model are slightly higher than the experimental data, but generally remain closer to them in comparison particularly with Z(5)-DD. 

\hspace{-0.2 cm} 2) In respect to Z(5)-H model, our results for the isotopes $^{128,130,132,134}$Xe are the same because the mass deformation parameter $a=0$. For $^{126}$Xe and $^{192,194,196}$Pt, all our results are slightly higher. Thus, the precision gained in terms of energy ratios, thanks to the introduction of a mass deformation parameter, is somewhat weakly lost in transition rates' outcomes. However, the precision of our model's calculations  persists versus Z(5)-DD. This latter observation is due to the flatness of the Hulth\'{e}n potential. Indeed, as it was mentioned in the introduction, the precision of transition rates calculations depends on this flatness insofar as it increases as the potential is flatter, when the $\beta$ coordinate increases too. From Fig.\ref{fig9}, one can see that Hulth\'{e}n potential is flatter than Davidson potential.

\begin{table*}
      	\begin{center}
      		\includegraphics[scale=0.85]{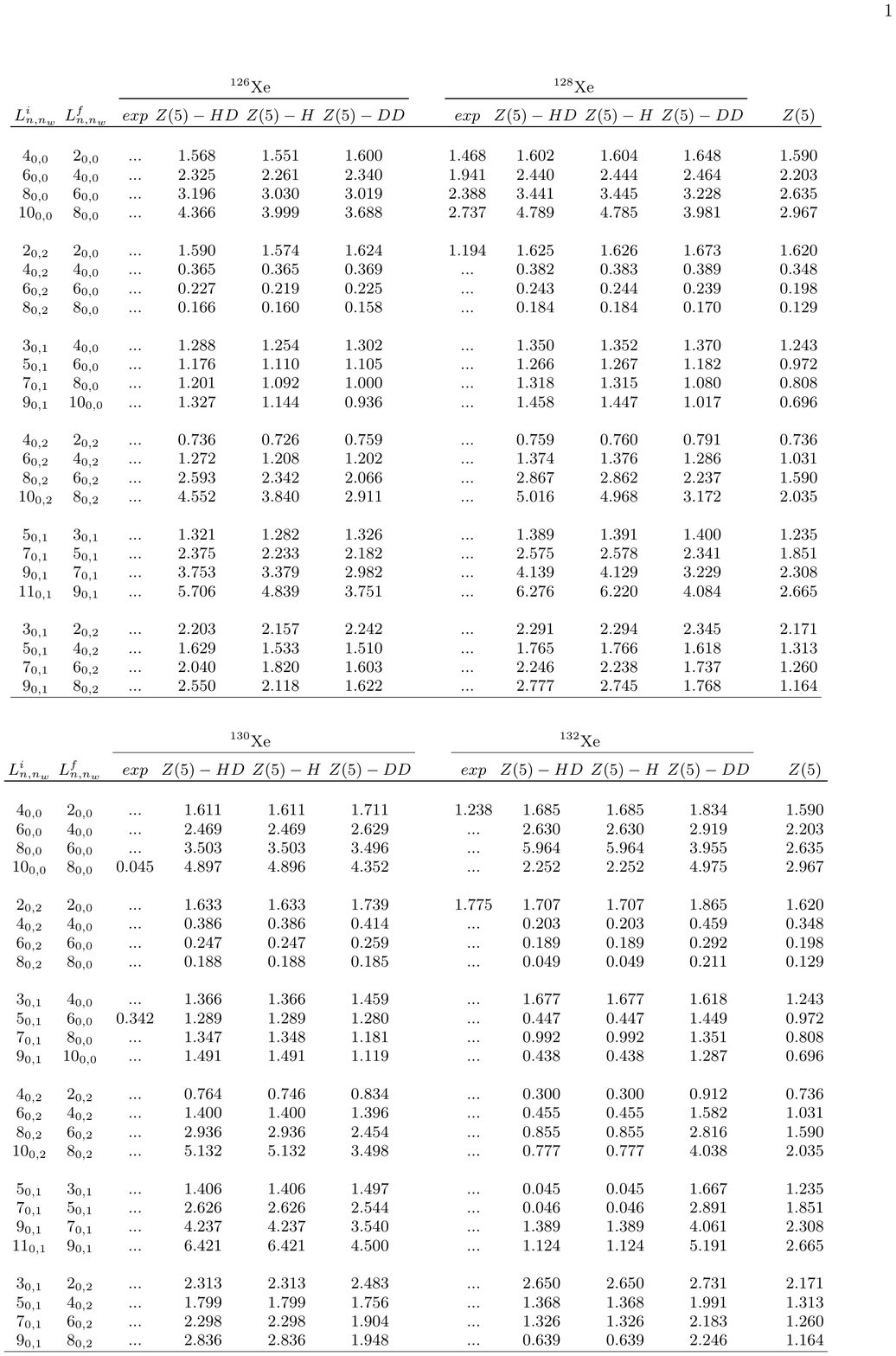}
      		\caption{The normalized $B(E2)$ transition rates of the Z(5)-HD model,  compared to the experimental data \cite{ww1}, Z(5)-H model \cite{Ch2}, Z(5)-DD model \cite{BO4} and  the free parameters model Z(5) \cite{Bo2} predictions for $^{126,128,130,132}$Xe isotopes.}
      \label{table:T4}		
      	\end{center}
      \end{table*}    

\begin{table*}
      	\begin{center}
      		\includegraphics[scale=0.85]{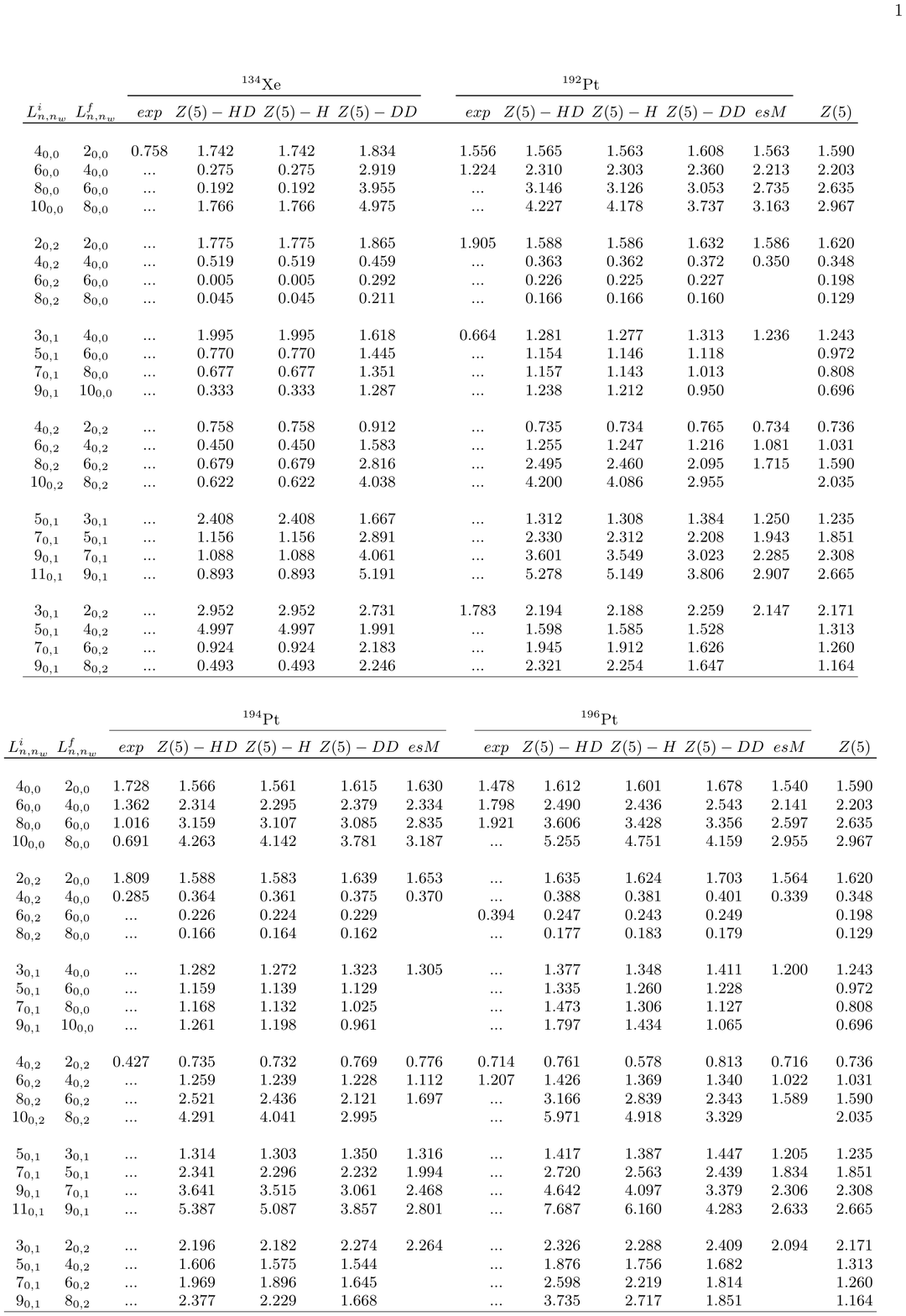}
         \caption{The normalized $B(E2)$ transition rates of the Z(5)-HD model,  compared to the experimental data \cite{ww1}, Z(5)-H model \cite{Ch2}, Z(5)-DD model \cite{BO4}, esM model \cite{Inc2014} and  the free parameters model Z(5) model \cite{Bo2} predictions for $^{134}$Xe , $^{192,194,196}$Pt isotopes.} 	
         \label{table:T5}	
      	\end{center}
      \end{table*}  
      
\begin{figure*}
      	\begin{center}
      		\includegraphics[scale=0.9]{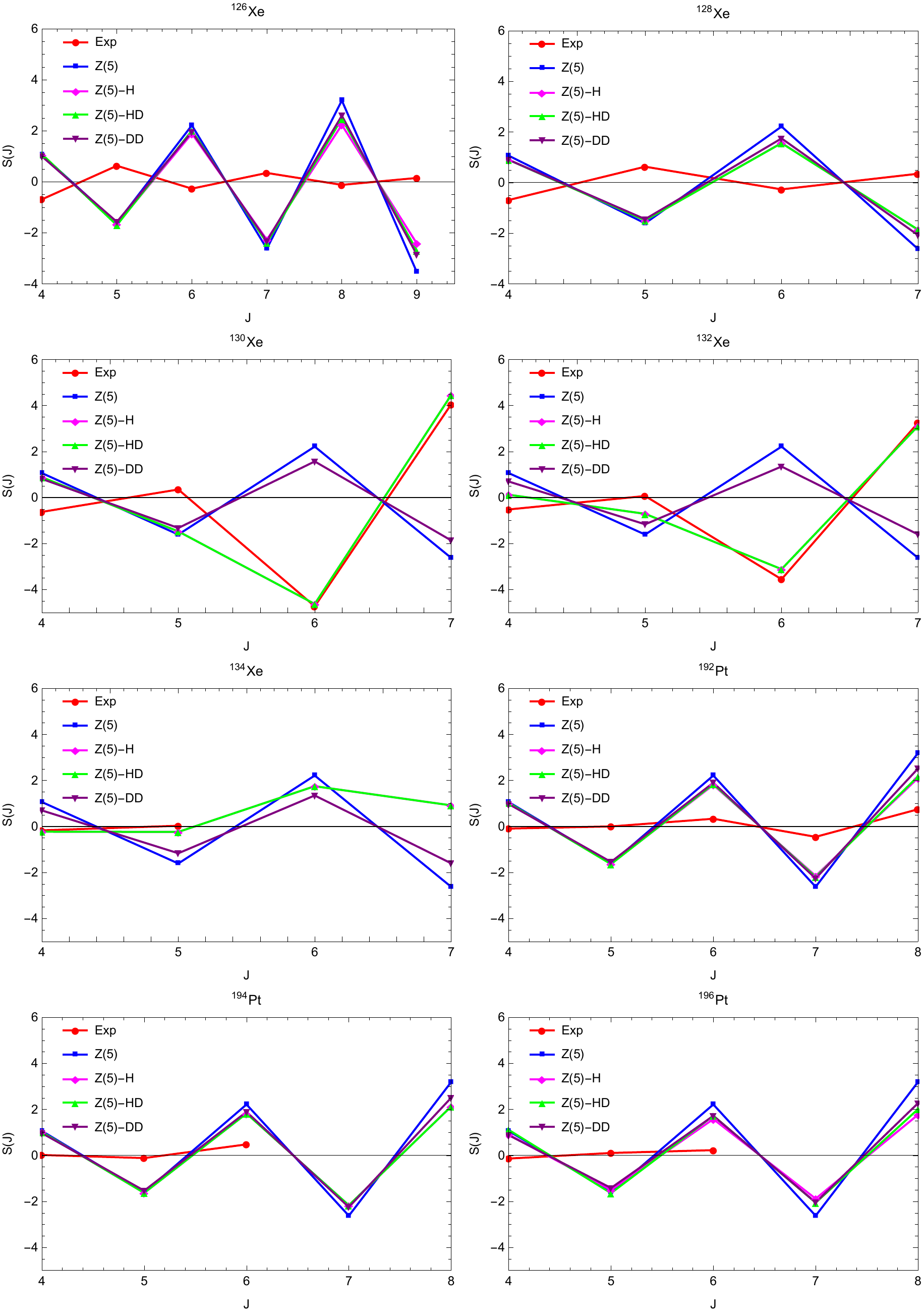}
      		\caption{The sensitive signature for triaxiality structure $S(J)$ ( Eq.\eqref{St} ) of isotopes $^{126,128,130,132,134}$Xe and $^{192,194,196}$Pt obtained in our calculation compared with experimental results \cite{ww1}, $Z(5)-H$ \cite{Ch2},  $Z(5)-DD$
      		 \cite{BO4}  and $Z(5)$ \cite{Bo2} models.}
      		\label{fig10}
      	\end{center}
      \end{figure*}

 Another important signature for triaxiality in nuclei  is the odd-even staggering of energy levels which happens in $\gamma$ band and which is described by the following relation \cite{Za1}:
\begin{equation}
S(J)=\frac{E(J_{\gamma}^{+})+ E((J-2)_{\gamma}^{+})-2\;E((J-1)_{\gamma}^{+})}{E(2_{1}^{+})}.
\label{St}                                                                  
\end{equation}
This relation gives the relative displacement of the $(J-1)_{\gamma}^{+}$ to the average of its neighbors, $J_{\gamma}^{+}$ and $(J-2)_{\gamma}^{+}$, normalized to the energy of the first excited state of the g.s band, $E(2_{1}^{+})$. It was shown \cite{Mc1} that $\gamma$-soft shapes exhibit staggering with negative $S(J)$ values at even-J and positive $S(J)$ values at odd-J spins. However, for triaxial nuclei the opposite signs are seen, i.e. positive $S(J)$ at even-J and negative $S(J)$ at odd-J. This is a sensitive probe of triaxiality, as shown, for example, in Ref. \cite{Mc1}. It should also be pointed out that for $^{126,128}$Xe the data show behavior opposite to all models (see Fig. \ref{fig10}).  The same holds more or less for $^{130,132}$Xe. In these two nuclei the Z(5)-HD model does get correctly the minimum at J=6, but its values show no staggering (jumping up and down). Actually occasional disagreements between theory and experiment can sometimes lead later to interesting physical insights, therefore they should be pointed out. Moreover, from Fig. \ref{fig10}, one can see that generally all studied nuclei exhibit a stronger odd-even staggering than that observed experimentally. However, one can remark that the amplitude of such an effect was attenuated within Z(5)-HD calculations tending to the experimental behavior. Besides, such an effect for $^{192}$Pt and $^{194}$Pt  is well reproduced by our model which concords with the previous studies in Refs. \cite{Mc1,Ch2}. So, the two isotopes $^{192}$Pt and $^{194}$Pt are considered as good candidates for triaxial deformation because of satisfying both signatures for triaxiality, namely: that of the triaxial rigid rotor and the staggering effect.
\section{Conclusion}
In the present work, we have solved the eigenvalues and eigenvectors problem with the Bohr collective Hamiltonian for triaxial nuclei within Deformation Dependent Mass formalism using  Hulth\'{e}n potential for  $\beta$-part and a new Ring-Shaped potential \cite{Ch2} for the $\gamma$ one. The obtained results with our proposed model dubbed Z(5)-HD were in overall agreement with the experimental data for energy ratios and transition rates of the nuclei $^{126,132,134}$Xe and $^{192,194,196}$Pt and comparatively better in general than other models. Moreover, our model was an improvement of the previously proposed one in Ref.\cite{Ch2}. Indeed, the introduction of a mass deformation parameter has significantly increased the precision of energy ratios calculations particularly. Besides, the flatness of Hulth\'{e}n potential versus Davidson one has played an important role in a satisfactory reproduction of experimental data of transition rates. Moreover, we  have shown that  Hulth\'{e}n potential is more appropriate for describing nuclei presenting a vibrational structure. Despite the difference between the parameters of the potential in both cases, namely: Hulth\'{e}n and Davidson, it was found a strong correlation between the values of the mass deformation parameter in these two cases which corroborates once again the fact that the mass deformation parameter is not just a simple one to be adjusted for reproducing experimental data, but should be considered as a model's structural one. Among the studied nuclei, we have confirmed that the better candidates for triaxiality were $^{192,196}$Pt. These two isotopes undergo both signatures for triaxiality, namely: that of the triaxial rigid rotor and the staggering effect. 


\begin{thebibliography}{99}
\bibitem{Von1983} O. von Roos, Phys. Rev. B 27, 7547 (1983).
\bibitem{Bas1}G. Bastard, Wave Mechanics Applied to Semiconductor Heterostructure (Les Editions de Physique, 1998).
\bibitem{Ser1}L. Serra and E. Lipparini, Europhys. Lett. 40, 667 (1997).
\bibitem{Saa1}F. A. de Saavedra, J. Boronat, A. Polls, and A. Fabrocini, Phys. Rev. B 50, 4248 (1994).
\bibitem{Pla1}A. R. Plastino, M. Casas, and A. Plastino, Phys. Lett. A 281, 297 (2001).
\bibitem{Alh1}A. D. Alhaidari, Int. J. Theor. Phys. 42,  2999 (2003).
\bibitem{Alh2}A. D. Alhaidari, Phys. Lett. A 322,  72 (2004).
\bibitem{Ch6}M. Chabab, A. Lahbas, M. Oulne, Phys. Rev. C 91, 064307(2015).
\bibitem{Bo8}D. Bonatsos, P. E. Georgoudis, N. Minkov, D. Petrellis, C. Quesne, Phys. Rev. C 88,  034316 (2013).
\bibitem{BO4}D. Bonatsos, P. E. Georgoudis, D. Lenis, N. Minkov and C. Quesne, Phys. Rev. C 83,  044321 (2011).
\bibitem{Soa}N. Soheibi, M. Eshghi and M. Bigdeli, Eur. Phys. J. Plus (2019) 134: 208.
\bibitem{Sob}N. Soheibi, M. Eshghi and M. Bigdeli, Eur. Phys. J. Plus 135, Article number: 75 (2020). 
\bibitem{Ch8} M. Chabab,  A. El Batoul, I. El-Ilali, A. Lahbas and M. Oulne, Eur. Phys. J. Plus 135, Article number: 201 (2020).
\bibitem{Ch9} M. Chabab,1 A. El Batoul,1 H. Hassanabadi, M. Oulne and S. Zare, Eur. Phys. J. Plus 131(11):387 (2016).

\bibitem{Jol10}R. V. Jolos and P. von Brentano, Phys. Rev. C 76, 024309 (2007).
\bibitem{Jol11}R. V. Jolos and P. von Brentano, Phys. Rev. C 77, 064317 (2008).
\bibitem{Jol12}R. V. Jolos and P. von Brentano, Phys. Rev. C 78, 064309 (2008).
\bibitem{Jol13}R. V. Jolos and P. von Brentano, Phys. Rev. C 79, 044310 (2009).
\bibitem{Jol14}R. V. Jolos and P. von Brentano, Phys. Rev. C 80, 034308 (2009).
\bibitem{Erm10}M. J. Ermamatov and P. R. Fraser, Phys. Rev. C 84, 044321 (2011).
\bibitem{Erm11}M. J. Ermamatov, P. C. Srivastava, P. R. Fraser, P. Stransky, and I. O. Morales, Phys. Rev. C 85, 034307 (2012). 

\bibitem{E5} F. Iachello, Physical Review Letters 85 (2000) 3580.
\bibitem{X5} F. Iachello, Phys. Rev. Lett. 87, 052502 (2001).
\bibitem{X3} D. Bonatsos, D. Lenis, D. Petrellis, P. Terziev, I. Yigitoglu, Physics Letters B 632 (2006) 238.
	\bibitem{Bo2} D. Bonatsos, D. Lenis, D. Petrellis, P.A. Terziev, Phys. Lett. B588, 172 (2004).
\bibitem{Z4}	D. Bonatsos, D. Lenis, D. Petrellis, P. Terziev, I. Yigitoglu, Physics Letters B 621 (2005) 102.
\bibitem{Ch7} M. Chabab, A. El Batoul, A. Lahbas, and M. Oulne, Nucl. Phys. A 953, 158-175 (2016). 
\bibitem{Ch2} M. Chabab, A. Lahbas, and M. Oulne, Eur. Phys. J. A, 51: 131 (2015).
\bibitem{BO6}  I. Yigitoglu and Dennis Bonatsos,Phys. Rev. C 83, 014303 (2011). 
  \bibitem{B1} A. Bohr, Mat. Fys. Medd. K. Dan. Vidensk. Selsk. 26, no. 14 (1952).
  
  \bibitem{Bag2005}  B. Bagchi, A. Banerjee, C. Quesne, and V. M. Tkachuk,
J. Phys. A: Math. Gen. 38, 2929 (2005).

  
  \bibitem{Fo2} L. Fortunato, Eur. Phys. J. A2 6 (s01), 1 (2005).
	\bibitem{Fo3} L. Fortunato, Phys. Rev. C 70, 011302 (2004).
	\bibitem{Fo4}  L. Fortunato, S. De Baerdemacker, and K. Heyde, Phys. Rev. C74, 014310 (2006).
	\bibitem{Bo3} D. Bonatsos, E. A. McCutchan, N. Minkov, R. F. Casten, P. Yotov, D. Lenis, D. Petrellis, I. Yigitoglu, Phys. Rev. C76, 064312 (2007).
	\bibitem{Radu} A. I. Budaca, R. Budaca, Eur. Phys. J. Plus 134, 145 (2019).  
	\bibitem{Hu1} L. Hulthén, Ark. Mat. Astron. Fys. A 28, 5 (1942).	
	\bibitem{Hu2} L. Hulthén, Ark. Mat. Astron. Fys. B29, 1 (1942).
	\bibitem{La1} U. Laha, C. Bhattacharyya, K. Roy, B. Talukdar, Phys. Rev. C38, 558 (1988).
	\bibitem{Ma1} P. Matthys, H. De Meyer, Phys. Rev. A38, 1168 (1988).
  \bibitem{Ji1} C.S. Jia, T. Chen, L.G. Cui, Phys. Lett. A373, 1621 (2009).
	\bibitem{Do1} S.H. Dong, W.C. Qiang, G.H. Sun, V.B. Bezerra, J. Phys.A: Math. Theor. 40, 10535 (2007).
	\bibitem{So1} A. Soylu, O. Bayrak, I. Boztosun, J. Phys. A: Math. Theor. 41, 065308 (2008).	
	\bibitem{Ci1}  H. Ciftci, R.L. Hall, N. Saad, J. Phys. A36, 11807 (2003).
	\bibitem{Ci2} H. Ciftci, R.L. Hall, N. Saad, J. Phys. Math. Gen. A38, 1147 (2005).
	\bibitem{Boz1} I. Boztosun, M. Karakoc, Chin. Phys. Lett. 24, 3028 (2007).
	\bibitem{Ci3} H. Ciftci, R.L. Hall, N. Saad, J. Phys. A36, 11807 (2003).
	\bibitem{Ik1} S.M. Ikhdair, R. Sever, J. Math. Chem. 42, 461 (2007).
	\bibitem{Ba2} O. Bayrak, G. Kocak, I. Boztosun, J. Phys. A: Math. Gen.39, 11521 (2006).
	\bibitem{Ag1} D. Agboola, Commun. Theor. Phys. 55, 972 (2011).
	\bibitem{Ch3} M. Chabab, A. Lahbas, M. Oulne, Int. J. Mod. Phys. E 21, 10 (2012).
	\bibitem{Fo1} L. Fortunato, Phys. Rev. C70, 011302 (2004).
	\bibitem{Bo1} D. Bonatsos, D. Lenis, N. Minkov, D. Petrellis, P.P. Raychev, P.A. Terziev, Phys. Lett. B584, 40 (2004).
	\bibitem{B2} A. Bohr, B.R. Mottelson, Nuclear Structure Vol. II: Nuclear
Deformations (Benjamin, New York, 1975).
  	\bibitem{Me} J. Meyer-ter-Vehn, Nucl. Phys. A249, 111 (1975).
	\bibitem{Int1} I.S. Gradshteyn and I.M. Ryzhik, Table of Integrals, Series, and Products (Seventh Edition)
	\bibitem{Ed1} A.R. Edmonds, Angular Momentum in Quantum Mechanics(Princeton University Press, Princeton, 1957).
	\bibitem{Ra} A.A. Raduta, P. Buganu, Phys. Rev. C83, 034313 (2011).
	\bibitem{Da} A.S. Davydov, G.F. Fillipov, Nucl. Phys. 8,237 (1958).
	\bibitem{Gr} W. Greiner, J.A. Maruhn, Nuclear Models (Springer, Berlin, 1996).
	\bibitem{Shar19}J. F. Sharpey-Schafer,  R. A. Bark,  S. P. Bvumbi,  T. R. S. Dinoko  and   S. N. T. Majola, Eur. Phys. J. A, 55 : 15 (2019).
	\bibitem{ww1} http://www.nndc.bnl.gov/nndc/ensdf/.	
     \bibitem{Ch5}M. Chabab, A. El Batoul, A. Lahbas, M. Oulne, J. Phys. G: Nucl. Part. Phys. 43, 12 (2016).	
	\bibitem{Za1}N.V. Zamfir, R.F. Casten, Phys. Lett. B 260, 265 (1991).
	\bibitem{Mc1} E.A. McCutchan, D. Bonatsos, N.V. Zamfir, R.F. Casten, Phys. Rev. C 76, 024306 (2007).	
	\bibitem{Inc2014} I. Inci, Int. J. Mod. Phys. E 23, 10 (2014).
		
\end{thebibliography}
\end{document}